\documentclass[aps,pra,onecolumn,notitlepage,superscriptaddress]{revtex4-1}
\usepackage[matrix,frame,arrow]{xy}
\usepackage{amsmath}

\newcommand{\qw}[1][-1]{\ar @{-} [0,#1]}

\newcommand{\gate}[1]{*{\xy *+<.6em>{#1};p\save+LU;+RU **\dir{-}\restore\save+RU;+RD **\dir{-}\restore\save+RD;+LD **\dir{-}\restore\POS+LD;+LU **\dir{-}\endxy} \qw}

\newcommand{\measureD}[1]{*{\xy*+=+<.5em>{\vphantom{\rule{0em}{.1em}#1}}*\cir{r_l};p\save*!R{#1} \restore\save+UC;+UC-<.5em,0em>*!R{\hphantom{#1}}+L **\dir{-} \restore\save+DC;+DC-<.5em,0em>*!R{\hphantom{#1}}+L **\dir{-} \restore\POS+UC-<.5em,0em>*!R{\hphantom{#1}}+L;+DC-<.5em,0em>*!R{\hphantom{#1}}+L **\dir{-} \endxy} \qw}

\newcommand{\multimeasureD}[2]{*+<1em,.9em>{\hphantom{#2}}\save[0,0].[#1,0];p\save !C *{#2},p+LU+<0em,0em>;+RU+<-.8em,0em> **\dir{-}\restore\save +LD;+LU **\dir{-}\restore\save +LD;+RD-<.8em,0em> **\dir{-} \restore\save +RD+<0em,.8em>;+RU-<0em,.8em> **\dir{-} \restore \POS !UR*!UR{\cir<.9em>{r_d}};!DR*!DR{\cir<.9em>{d_l}}\restore \qw}

\newcommand{\multigate}[2]{*+<1em,.9em>{\hphantom{#2}} \qw \POS[0,0].[#1,0];p !C *{#2},p \save+LU;+RU **\dir{-}\restore\save+RU;+RD **\dir{-}\restore\save+RD;+LD **\dir{-}\restore\save+LD;+LU **\dir{-}\restore}
\newcommand{\ghost}[1]{*+<1em,.9em>{\hphantom{#1}} \qw}

\newcommand{\Qcircuit}[1][0em]{\xymatrix @*[o] @*=<#1>}  
 \renewcommand{\Qcircuit}[1][0em]{\xymatrix @*=<#1>}

\newcommand{\pureghost}[1]{*+<1em,.9em>{\hphantom{#1}}}
\newcommand{\multiprepareC}[2]{*+<1em,.9em>{\hphantom{#2}}\save[0,0].[#1,0];p\save !C
  *{#2},p+RU+<0em,0em>;+LU+<+.8em,0em> **\dir{-}\restore\save +RD;+RU **\dir{-}\restore\save
  +RD;+LD+<.8em,0em> **\dir{-} \restore\save +LD+<0em,.8em>;+LU-<0em,.8em> **\dir{-} \restore \POS
  !UL*!UL{\cir<.9em>{u_r}};!DL*!DL{\cir<.9em>{l_u}}\restore}
\newcommand{\prepareC}[1]{*{\xy*+=+<.5em>{\vphantom{#1\rule{0em}{.1em}}}*\cir{l^r};p\save*!L{#1} \restore\save+UC;+UC+<.5em,0em>*!L{\hphantom{#1}}+R **\dir{-} \restore\save+DC;+DC+<.5em,0em>*!L{\hphantom{#1}}+R **\dir{-} \restore\POS+UC+<.5em,0em>*!L{\hphantom{#1}}+R;+DC+<.5em,0em>*!L{\hphantom{#1}}+R **\dir{-} \endxy}}
\newcommand{\poloFantasmaCn}[1]{{{}^{#1}_{\phantom{#1}}}}

\usepackage{graphicx,color}
\usepackage{dcolumn}
\usepackage{bm}
\usepackage{amsmath,amssymb,mathrsfs}
\usepackage{url}
\usepackage{hyperref}

\newcommand{\R}{\mathbb{R}}

\newcommand{\set}[1]{\mathsf{#1}}

\newcommand{\Span}{{\mathsf{Span}}}

\def\>{\rangle}
\def\<{\langle}

\newcommand{\su}[1]{\mathbf{#1}}
\newcommand{\bs}[1]{\boldsymbol{#1}}

\newcommand{\map}[1]{\mathcal{#1}}

\newcommand{\Sys}{{\mathsf{Sys}}}
\newcommand{\St}{{\mathsf{St}}}
\newcommand{\Eff}{{\mathsf{Eff}}}
\newcommand{\Pur}{{\mathsf{Pur}}}
\newcommand{\Det}{{\mathsf{Det}}}
\newcommand{\Transf}{{\mathsf{Transf}}}

\newcommand{\Ext}{{\mathsf{Ext}}}

\def\rA{{\rm A}}
\def\rB{{\rm B}}
\def\rC{{\rm C}}
 
\def\rE{{\rm E}} 
\def\rF{{\rm F}}

\def\rI{{\rm I}}
\def\rR{{\rm R}} 
\def\rS{{\rm S}}

\newtheorem{ass}{Assumption}

\newtheorem{prin}{Principle}
\newtheorem{theo}{Theorem}
\newtheorem{lem}{Lemma}
\newtheorem{prop}{Proposition}
\newtheorem{cor}{Corollary}
\newtheorem{defi}{Definition}
\newtheorem{rem}{Remark}

\def\Proof{{\bf Proof.~}}
\def\qed{$\blacksquare$ \newline}

\begin{document}

\title{Process tomography in general probabilistic theories}
\author{Giulio Chiribella}
\affiliation{QICI Quantum Information and Computation Initiative, Department of Computer Science, The University of Hong Kong}
\affiliation{Department of Computer Science, University of Oxford, Wolfson Building, Parks Road, Oxford, UK} 
\affiliation{HKU-Oxford Joint Laboratory for Quantum Information and Computation}
\affiliation{Perimeter Institute for Theoretical Physics, 31 Caroline Street North, Waterloo, Ontario, Canada}

\begin{abstract}
Process tomography, the  experimental characterization of physical processes,  is a central task in science and engineering.
 Here we investigate   the  axiomatic requirements that guarantee the in-principle feasibility of process tomography    in  general physical theories.  
Specifically,  we explore the requirement that process tomography  should be achievable with a finite number of auxiliary systems and with a finite number of input states.  We show that this requirement is satisfied   
   in every theory equipped with  universal extensions,  that is, correlated states from which all other correlations can be generated locally with non-zero probability.  
     We show that universal extensions are guaranteed to exist in two cases: (1)  theories permitting conclusive state teleportation, and  (2)  theories  satisfying three  properties  of Causality,  Pure Product States, and  Purification.   In case (2), the existence of universal extensions follows from a symmetry property of Purification, whereby all pure bipartite states with the same marginal on one system are locally interconvertible.  
    Crucially, our results hold even in theories that do not satisfy  Local Tomography, the property that the state of any composite system can be identified from  the correlations of local measurements.  
    Summarizing,  the existence of universal extensions, without any additional requirement of Local Tomography,   is a sufficient guarantee for the characterizability of physical processes using a finite number of auxiliary systems and a finite number of input states. 
\end{abstract}

\maketitle

\section{Introduction} 
The experimental characterization  of  physical processes is an important component of the scientific method.   
Such a characterization, known as process tomography,  is  widely adopted in  classical   \cite{beck2012process} and quantum technologies \cite{chuang1997prescription,poyatos1997complete,leung2000towards,leung2003choi,d2001quantum,dur2001nonlocal,altepeter2003ancilla,mohseni2008quantum,lobino2008complete,scott2008optimizing,bisio2009optimal,shabani2011efficient,baldwin2014quantum}.   
 In general, one can regard the in-principle feasibility of process tomography as a requirement for  the construction of new physical theories:   arguably,   a sensible physical theory  should  describe  processes that---{at least in principle}---can be characterized  experimentally.   
 
  Operational axioms  inspired by  process tomography were first proposed by D'Ariano  in a sequence of works \cite{dariano2006missing,dariano2006how,dariano2007operational,maurobook} where they featured as potential candidates for an axiomatization of quantum theory.     In this paper, we will explore the conditions that guarantee the feasibility of  process tomography in general physical theories, independently of the quantum axiomatization problem.    Specifically, our goal will be to identify physical conditions that guarantee  
   the achievability  of process tomography using a finite number of auxiliary systems and with a finite number of input states.    

 To understand the role of auxiliary systems in process tomography, it is useful to consider first a popular approach known as {\em standard process tomography} \cite{chuang1997prescription,poyatos1997complete}, where no auxiliary system is used.   Standard process tomography  aims at characterizing processes by letting them act on a  set of input states and by analyzing the output with a set of measurements, as in the following schematic
\begin{align}\label{localtest}
\begin{aligned}
 \Qcircuit @C=1em @R=.7em @!R { & \prepareC{\rho^{(i)}} & \qw \poloFantasmaCn{\rA} &  \gate{\map P^{\phantom{j}}}  &  \qw \poloFantasmaCn{\rB} & \measureD{\su m^{(j)}} }
\end{aligned} ~~.
\end{align}
Here,  an unknown  process $\map P$ (with input system $\rA$ and output system $\rB$)  is tested on a {
 set of fiducial states  $\{\rho^{(i)}\}_{i  =1}^N$ with a set of fiducial measurements $\{ \su m^{(j)}\}_{j=  1}^{M}$. We use the boldface notation $\su  m^{(j)}$ to indicate  that the $j$-th measurement has multiple outcomes, labelled by some index that is not written down explicitly.}  
By choosing a complete set of fiducial states and measurements, one can then identify the action of the process  $\map P$ on its input system---that is, one can uniquely determine   the function $f_{\map P}$ that maps a generic input state of system $\rA$   into the corresponding output state of system $\rB$.


In classical  and quantum physics,  standard process tomography is sufficient to completely characterize physical processes:  
  by determining the function $f_{\map P}$, one can identify  the action  of the process $\map P$ in every possible experiment, { even including experiments where the process $\map P$ acts locally on a part of a composite system.    For example, this includes experiments of the form  
     \begin{align}\label{ancillaassisted}
\begin{aligned}
 \Qcircuit @C=1em @R=.7em @!R { & \multiprepareC{1}{\Psi} & \qw \poloFantasmaCn{\rA} &  \gate{\map P}  &  \qw \poloFantasmaCn{\rB} & \multimeasureD{1}{{\su M}}   
 \\
  &\pureghost{\Psi }     &  \qw  \poloFantasmaCn{\rC}  &  \qw  &  \qw  &  \ghost{{\su M}} 
  }
\end{aligned} \quad  ,
\end{align}
where  $\map P$ is applied locally on a joint state $\Psi$ of systems $\rA$ and system $\rC$, and a joint measurement $\su M$ is performed on  systems $\rB$  and  $\rC$. }   

  In classical and quantum theory,   the outcome probabilities of all experiments of type (\ref{ancillaassisted}) are uniquely determined by the outcome probabilities of experiments of type~\eqref{localtest}.
  The origin of this favourable situation is a property,  known as {\em Local Tomography}, whereby   the states of every composite system can be uniquely identified by performing local measurements on the components \cite{araki1980characterization,wootters1990local,hardy2001quantum,dariano2006how,dariano2007operational,barrett2007information,barnum2007generalized,chiribella2010probabilistic,hardy2011reformulating,chiribella2012quantum,dariano2017quantum}.   {  In quantum foundations, Local Tomography has often been taken as an axiom for the characterization of standard quantum theory (on complex Hilbert spaces)\cite{hardy2001quantum,chiribella2011informational,dakic2011quantum,masanes2011derivation,hardy2011reformulating,masanes2012existence,barnum2014higher,wilce2019conjugates,selby2021reconstructing} and as a principle for the construction of new physical theories \cite{barrett2007information,barnum2007generalized,chiribella2010probabilistic}.       In locally tomographic theories,}  the action of a process on its input system uniquely determines the action of the process on every   composite system 
\cite{chiribella2010probabilistic}.  As a result,  standard process tomography provides a complete characterization of physical processes.

The situation is radically different when   Local Tomography  does not hold.  A simple counterexample arises in real-vector-space quantum theory  \cite{stueckelberg1960quantum}, a variant of standard  quantum theory  that violates Local Tomography  \cite{araki1980characterization,wootters1990local} and  deviates from standard quantum theory in a number  of operational tasks \cite{wootters1990local,chiribella2010probabilistic,wootters2012entanglement,hardy2012limited,aleksandrova2013real,chiribella2013quantum,wootters2014rebit,wootters2016optimal}.  In real-vector-space quantum theory,   one can find explicit examples of processes that have the same input-output function, and yet act in  completely different ways on composite systems  \cite{chiribella2016quantum}.    

When Local Tomography  fails,  the  only way to unambiguously identify a process  is  to  test its action on composite systems, using an approach known as {\em ancilla-assisted process tomography} \cite{leung2000towards,d2001quantum,dur2001nonlocal,leung2003choi,altepeter2003ancilla}.   This approach consists in performing experiments of the type of Eq. (\ref{ancillaassisted}). 
  Now,  the  crucial question is:  which  auxiliary systems $\rC$   have to be taken into account?   In the worst case,  the answer could be {\em ``all  possible systems''}:  the  characterization of an unknown  physical process may require  experiments performed, as it were, on the whole universe.  
    It seems natural to demand, as a basic principle,  that physical theories should be free from this  pathology. { In  a sensible physical theory, it should be in principle possible to identify  any  given process up to any desired level of accuracy using a finite number of experimental settings. This means: testing the process with a finite number of auxiliary  systems, preparing a finite number of states, and performing a finite number of measurements, each with a finite number of outcomes.  
   
   In this paper, we  will focus on the requirements of a finite number  of auxiliary systems/finite number of inputs states, and we will consider the case of process tomography with ideally  perfect accuracy.  
      We will require that, at least in principle, physical processes can be completely characterized by probing their action with a finite number of auxiliary systems.    Note that, in turn,  every {\em finite} set of  auxiliary systems  $\{\rC_i\}_{i=1}^k$  can be replaced without loss of generality by a {\em single} auxiliary system  $\rR$, for example by merging all the systems of the finite  set into a composite system \footnote{Most of the frameworks for  general probabilistic theories include a notion of system composition, denoted by $\otimes$ and corresponding to a generalization of the tensor product in quantum theory \cite{hardy2001quantum,barnum2007generalized,barrett2007information,chiribella2010probabilistic,chiribella2011informational,barnum2011information,hardy2011foliable,hardy2013formalism,chiribella2014dilation,chiribella2016quantum,hardy2016reconstructing,dariano2017quantum,gogioso2018categorical,selby2021reconstructing}.   As long as the number of components is finite, the composite system is well defined in all these frameworks.    At the mathematical level, the composition $\otimes$ is defined as the tensor product in a monoidal category describing physical systems and processes between them \cite{abramsky2004,coecke2006kindergarten,abramsky2008,coecke2010quantum}.    When monoidal categories are used as a framework for general physical theories, they are often called {\em process theories} \cite{coecke2018picturing}.      }.   This observation motivates the following requirement:
 \begin{prin}[Dynamically Faithful Systems]\label{prin:feasibility}  For every pair of systems $\rA$ and $\rB$, there exists an auxiliary system $\rR$     such that all processes with input $\rA$ and output $\rB$ can be completely characterized by their action on 
the 
states of the composite system $\rA\otimes \rR$.
\end{prin}
We will also consider a slightly stronger requirement, inspired by D'Ariano's early axiomatization works  \cite{dariano2006missing,dariano2006how,dariano2007operational,maurobook}.     The requirement  stipulates  the achievability of process tomography with a {\em single} state of a suitable composite system.     \begin{prin}[Dynamically Faithful States]\label{prin:dynfaith1}  For every pair of systems $\rA$ and $\rB$, there exists an auxiliary system $\rR$     such that all processes with input $\rA$ and output $\rB$ can be completely characterized by their action on a single state of the composite system $\rA\otimes \rR$.
\end{prin}

 }

Dynamically Faithful Systems/Dynamically Faithful  States are  strictly weaker requirements than Local Tomography. For example, real-vector-space quantum theory violates Local Tomography  but satisfies Dynamically Faithful States (and therefore Dynamically  Faithful Systems):   every real-vector-space process with input $\rA$  can be uniquely  characterized by its action on an entangled state of  the composite system $\rA \otimes \rA$  \cite{chiribella2010probabilistic}.    It is also worth noting that  Dynamically Faithful Systems/Dynamically Faithful  States  do not imply finite dimensionality:  for example, all the processes on a quantum system with separable Hilbert space can be characterized by preparing a single bipartite state \cite{d2001quantum}.    {    Nevertheless, in this paper we will be mostly concerned with finite dimensional systems for simplicity of presentation  \footnote{Note that we do not make any assumption on the size (however defined) of the system $\rR$.  A further strengthening of our requirements would by to demand that   the size of system $\rR$  be bounded in terms of the sizes of systems $\rA$ and $\rB$.  For example, one could demand that  $\rR$  be embeddable into $\rA \otimes \rB$, or that $\rR$ be  operationally equivalent to system $\rA$, as it happens in quantum theory and in classical probability theory.  }. }

The core result of the paper is a mathematical characterization of the   theories satisfying Dynamically Faithful States. The characterization  is valid for a broad class of physical theories---so broad that, in fact,  it even includes  non-causal theories where the choices of experiments performed in the future can affect the outcome probabilities of experiments performed in the past.   After providing the mathematical characterization, we identify physical conditions that guarantee Dynamically Faithful States.      Specifically, we show that  Dynamically Faithful States holds in all theories where   all  correlations  of a given physical system with its environment  can be probabilistically generated by local operations on  a single ``universally correlated state.''     We call this property Universal Extension.  
  Two important  classes of theories satisfying Universal Extension are  {\em (1)} the theories allowing for  conclusive  teleportation \cite{abramsky2004,barnum2012teleportation}, and  {\em (2)} the theories satisfying  three requirements of Causality,  Pure Product States, and Purification.  Informally, Causality is the requirement that   the outcome probabilities of present experiments are independent on the choice of future experiments   \cite{chiribella2010probabilistic,coecke2010causal,chiribella2011informational,chiribella2012quantum,coecke2013causal,chiribella2014dilation,chiribella2016quantum,dariano2017quantum,coecke2018picturing}.  Pure Product States is the requirement that if the parts of a composite system are in pure states, then the whole composite system is in a pure state. Purification is   the requirement  that every mixed state can be extended to a pure state, { and that  such extension is unique up to  local symmetry transformations \cite{chiribella2010probabilistic,chiribella2011informational,chiribella2012quantum,chiribella2013quantum_route,chiribella2014dilation,chiribella2015conservation,chiribella2016quantum,dariano2017quantum}.  Remarkably, the symmetry property of purifications guarantees that every purification is a universal extension, and therefore every theory with purification has  Dynamically Faithful States.  } Our main results can be summarized by the following logical  implications: 
  \begin{enumerate}
   \item Conclusive Teleportation   $\Longrightarrow$ Universal Extension  
  \item Causality, Pure Product States, and Purification  $\Longrightarrow$ Universal Extension
  \item  Universal Extension  $\Longrightarrow$  Dynamically Faithful States  $\Longrightarrow$  Dynamically Faithful Systems.
  \end{enumerate}
The overall conclusion  of our results is that Universal Extension  guarantees the in-principle characterizability of physical processes, without invoking any assumption of Local Tomography.

\section{Materials and Methods}   

In this section we briefly introduce the framework and the notation used in the paper.  

\subsection{Operational-probabilistic theories} 

A major  trend in   quantum foundations is the study of information-processing tasks in a broad class of general probabilistic theories, which  include quantum theory as a special case \cite{hardy2001quantum,barnum2007generalized,barrett2007information,chiribella2010probabilistic,chiribella2011informational,barnum2011information,hardy2011foliable,hardy2013formalism,chiribella2014dilation,chiribella2016quantum,hardy2016reconstructing,dariano2017quantum}.      
  A convenient approach to general probabilistic theories is provided by  the framework of {\em operational-probabilistic theories (OPTs)}  \cite{chiribella2010probabilistic,chiribella2011informational,hardy2011foliable,hardy2013formalism,chiribella2014dilation,chiribella2016quantum,hardy2016reconstructing,dariano2017quantum,tull2016operational,gogioso2018categorical}.  The framework consists of  two distinct conceptual ingredients: an operational structure, describing circuits that produce outcomes, and a probabilistic structure, which assigns probabilities to the outcomes.
  The operational structure is inspired by    the approach  of categorical quantum mechanics  \cite{abramsky2004,coecke2006kindergarten,abramsky2008,coecke2010quantum} (see also \cite{coecke2018picturing} for a recent presentation), and follows it  rather closely, although  there are a few relevant  differences in the way classical outcomes are treated \cite{tull2016operational,gogioso2018categorical}.   

The OPT framework  describes a set of  experiments that can be performed on a given set of systems with a given set of physical processes. The framework is based on a primitive notion of composition, whereby every pair of physical systems $\rA$ and $\rB$ can be combined into a {\em composite system} $\rA\otimes \rB$.  The set of all physical systems is closed under composition, and will be denoted by $\Sys$. Physical processes can be combined in sequence or in parallel to build
circuits, such as
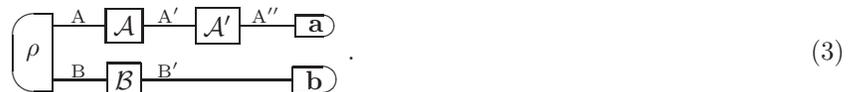
\begin{align}\label{circuit}
\begin{aligned}\Qcircuit @C=1em @R=.7em @!R { & \multiprepareC{1}{\rho} & \qw \poloFantasmaCn{\rA} & \gate{\map A} & \qw \poloFantasmaCn{\rA'} & \gate{\map A'} & \qw \poloFantasmaCn{\rA''} &\measureD{\su a} \\ & \pureghost{\rho} & \qw \poloFantasmaCn{\rB} & \gate{\map B} & \qw \poloFantasmaCn{\rB'} &\qw &\qw &\measureD{ \su b} }\end{aligned}~.
\end{align}
In this example, $\mathrm{A}$, $\mathrm{A}', \rA'', \rB$, and $\rB'$ are \emph{systems}, $\rho$
is a  \emph{state} of the  composite system $\rA\otimes \rB$, $\mathcal{A}$, $\mathcal{A}'$ and $\mathcal{B}$
are \emph{transformations} (a.k.a. {\em processes}), $\su a$ and $\su b$ are \emph{measurements}.   We use  boldface fonts $\su a$ and $\su b$ for measurements in order to indicate that they generally  have multiple  outcomes. When necessary, we will explicitly write  $\su a =  (a_x)_{x\in\set X}$ and $\su b  =  (b_y)_{y\in\set Y}$,  where $x$ and $y$ are measurement outcomes,  and $\set X$ and $\set Y$ are  the sets of possible outcomes of the two measurements, respectively.     
Here $a_x$ and $b_y$ play the same role as the linear operators associated to the outcomes of quantum measurements.  Following a traditional terminology dating back to   Ludwig  \cite{ludwig1985foundations}, we will call them  {\em effects}.  

Among the physical systems, every OPT includes a trivial system, denoted by $\mathrm{I}$ and  corresponding to the degrees
of freedom ignored by  theory. 
States (resp.\ effects) are transformations with the trivial system
as input (resp.\ output). 

Transformations from the trivial system to itself are called {\em scalars}.  Physically, they are associated to  circuits with no external wires, like the
circuit in Eq.~\eqref{circuit}.   In the OPT framework, scalars are typically identified with   numerical probabilities, valued in the interval $[0,1] \subset \R$.    More generally, other identifications are also possible, including  signed probabilities or possibilities, as shown in the work of Gogioso and Scandolo \cite{gogioso2018categorical}. { In the following, we will not make any assumption on the nature of the scalars, although we will often call them ``probabilities'' to facilitate the connection to the existing literature in quantum foundations.    }     

%

For generic systems $\rA$ and $\rB$, we denote
by 
\begin{itemize}
\item $\Transf (\rA\to \rB)$ the set of transformations
from $\rA$ to $\rB$, also called transformations {\em  of type  $\rA\to \rB$}
\item $\St\left(\rA \right)  : =  \Transf(\rI\to \rA)$ the set of states of system
$\mathrm{A}$,
\item $\Eff (\rA):  = \Transf(\rA\to \rI)$ the set of effects on system $\mathrm{A}$,
\item   $\mathcal B \circ \mathcal A$   (or $\mathcal B \mathcal A$, for short) the sequential composition of two transformations $\mathcal A$ and $\mathcal B$, with the input of $\mathcal B$ matching the output of $\mathcal A$,
\item $\mathcal{A}\otimes\mathcal{B}$ the parallel composition  of the transformations $\mathcal{A}$ and $\mathcal{B}$.
\item  $\map I_{\rA}$  the identity transformation on system $\rA$
{
 \item $1:  =  \map I_{\rI}$ the identity transformation on the trivial system.}  
\end{itemize}

\begin{defi}\label{def:opind} 
We say that two transformations  $\map P$ and $\map P'$  are {\em operationally indistinguishable} \cite{chiribella2010probabilistic} if they  give rise to the same probabilities in all possible experiments,  namely 
\begin{align}\label{opindistinguishable}
\begin{array}{c} \begin{aligned} \Qcircuit @C=1em @R=.7em @!R {  & \multiprepareC{1}{\Gamma} & \qw \poloFantasmaCn{\rA} &  \gate{\map P}  &  \qw \poloFantasmaCn{\rB} & \multimeasureD{1}{E}  \\  
	 & \pureghost{\Gamma} & \qw \poloFantasmaCn{\rC} &  \qw &\qw     &\ghost{E}} \end{aligned} \quad   =   
   \begin{aligned} \Qcircuit @C=1em @R=.7em @!R { & \multiprepareC{1}{\Gamma} & \qw \poloFantasmaCn{\rA} &  \gate{\map P'}  &  \qw \poloFantasmaCn{\rB} & \multimeasureD{1}{E}   &\qquad &\qquad & \\ 
	 & \pureghost{\Gamma} & \qw \poloFantasmaCn{\rC} &  \qw &\qw   &  \ghost{E}  &  \qquad  & \qquad &  \qquad } \end{aligned}	   \begin{array}{c} \forall \rC \in \Sys  \\  \forall \Gamma  \in \St (\rA\otimes \rC)\\
	 \forall  E \in \Eff (\rB\otimes \rC)  \end{array}   \end{array}
	 \end{align}  
	 \end{defi}
In this paper we will focus our attention on {\em quotient theories} \cite{chiribella2014dilation,chiribella2016quantum}  where physical transformations that are operationally indistinguishable are identified: in other words, we will assume that  condition   (\ref{opindistinguishable}) 
 implies the equality 
	 \begin{align}
\begin{aligned} 
\Qcircuit @C=1em @R=.7em @!R {  & \qw \poloFantasmaCn{\rA} &  \gate{\map P}  &  \qw \poloFantasmaCn{\rB} & \qw}
\end{aligned}  
\quad =  
\begin{aligned} 
\Qcircuit @C=1em @R=.7em @!R {  & \qw \poloFantasmaCn{\rA} &  \gate{\map P'}  &  \qw \poloFantasmaCn{\rB} & \qw}
\end{aligned}~.
\end{align}	
When the scalars are positive real numbers, the set of transformations $\Transf  (\rA\to \rB)$ in the quotient theory can be regarded as elements of  an ordered vector space \cite{chiribella2010probabilistic,chiribella2014dilation,chiribella2016quantum,schmid2020unscrambling}.

Every transformation takes place in  a {\em test}, that is,  a non-deterministic process that generates an outcome out of a set of possible  outcomes.   Mathematically, a test  of type $\rA\to \rB$, with outcomes in $\set X$,  is a list of transformations of type $\rA\to \rB$ indexed by  elements of the outcome set $\set X$. We denote such a list by $\bs{ \map T}  =  (\map T_x)_{x\in\set X}$, with $\map T_x\in\Transf(\rA\to \rB)$ for  every $x\in\set X$. 

Tests of type $\rI\to \rA$ are called {\em preparation tests}, or {\em sources}.   A preparation test ${\bs \rho}  =  (\rho_x)_{x\in\set X}$ describes the non-deterministic preparation \footnote{In quantum theory, the states in a preparation test are {\em subnormalized} density matrices,  and the trace of each subnormalized density matrix is interpreted as the probability of the corresponding preparation. } of the state $\rho_x$, heralded by an outcome $x\in\set X$.   
Tests of type $\rA\to \rI$ are called {\em measurements}. A measurement  $\su a  =  (a_x)_{x\in\set X}$ is a collection of effects labelled by measurement outcomes.

A transformation is  {\em deterministic}  if it is part of a test with a {\em single outcome}. By definition, deterministic transformations can be performed deterministically by setting up the corresponding test.   We use the notations  
\begin{itemize}
\item $\Det \Transf (\rA\to \rB)$ for the set of deterministic transformations of type $\rA\to \rB$,
\item $\Det\St (\rA)$ for the set of deterministic states of system $\rA$,
\item   $\Det\Eff (\rA)$ for the deterministic effects of system $\rA$. 
\end{itemize}            
For the trivial system $\rI$, we assume that there exists only one deterministic transformation of type $\rI\to \rI$, namely the identity transformation $1$. This assumption was introduced by Coecke in Ref. \cite{coecke2016terminality}, motivated by  the  interpretation of the scalars as ``probabilities,'' which suggests that there should be only one deterministic scalar, corresponding to the notion of certainty. 
 
We assume that the set of tests is closed under {\em coarse-graining}, the operation of joining together two or more outcomes.   Given a test  $\bs {\map T} =  (\map T_x)_{x\in\set X}$ and a partition of the outcome set $\set X$ into disjoint subsets $(\set X_y)_{y\in\set Y}$, we define the {\em coarse-grained test}  $\bs {\map  S}  = (\map S_y)_{y\in\set Y}$  as the test with transformations 
\begin{align}
\map S_y  :  =  \sum_{x\in\set X_y}  \, \map T_x \, .
\end{align}

At this level, the symbol of sum is just a notation for the operation of coarse-graining:   { the only mathematical requirements on the coarse-graining operation is that it distributes over parallel and sequential composition, namely  
\begin{align}\label{coarse1}
\map A  \otimes  \left(   \sum_{x\in\set X_y}  \, \map T_x\right)    =    \sum_{x\in\set X_y}  \,\left(\map A\otimes \map T_x\right)    \qquad {\rm and}  \qquad   \left(   \sum_{x\in\set X_y}  \, \map T_x\right) \otimes \map A     =    \sum_{x\in\set X_y}  \,\left( \map T_x \otimes \map A \right)   \, , \end{align}
for every transformation $\map A$,  
\begin{align}\label{coarse2}
\map B  \circ  \left(   \sum_{x\in\set X_y}  \, \map T_x\right)    =    \sum_{x\in\set X_y}  \,\left(\map B \circ  \map T_x\right)   \, , \end{align}
for every transformation $\map B$ with input matching the output of  $\bs {\map T}$, and 
\begin{align}\label{coarse3}
 \left(   \sum_{x\in\set X_y}  \, \map T_x\right)   \circ \map C =    \sum_{x\in\set X_y}  \,\left(  \map T_x \circ \map C \right)   \, , \end{align}
for every transformation $\map C$ with output matching the input of $\bs {\map T}$.   Closely related notions of coarse-graining were proposed  in the works of Tull \cite{tull2016operational}, Gogioso and Scandolo \cite{gogioso2018categorical}, where Eqs.  (\ref{coarse1}-\ref{coarse3}) were assumed together with a few additional requirements.    

Since  there is  only one deterministic transformation of type $\rI\to \rI$ (the scalar 1), every test $\su p  =  (p_x)_{x\in\set X}$ of type $\rI \to \rI$ satisfies  the  normalization condition 
\begin{align}
\sum_{x\in\set X} p_x    =  1   \, .
\end{align} 
This condition is the analogue of the normalization of a probability distribution.  For this reason, in the following we will call the tests of type $I\to I$  ``probability distributions,'' even though in general the scalars may  not be  real numbers in the interval $[0,1]$. 

When the scalars are real numbers, the  sum notation used above is consistent with the notion of sum in the ordered  vector space  containing  the physical transformations  in the quotient  theory  \cite{chiribella2010probabilistic,chiribella2014dilation,chiribella2016quantum,schmid2020unscrambling} (cf. Definition \ref{def:opind} and comments below it).  }

\subsection{Framework assumptions}
  
 In this paper we will make four basic assumptions { that are shared by most probabilistic theories  in the literature.   We spell out the assumptions explicitly because they will play a significant role in our results, and it is convenient  to keep track of which assumption is used in which result.   For example, we will write ``Lemma 1 (A1,A2, A3)'' to state that that Lemma 1 follows from assumptions 1, 2, and 3. 
 }    
 
 Our first assumption is that all non-deterministic tests arise from deterministic processes  followed by measurements.    This assumption is compatible with the idea  that the outcomes can be read-out from some physical system, like the display of a device, and that such system is also described by the theory.   The readout process is then realized as a deterministic transformation, followed by a measurement on the display: 
\begin{ass}[Displays \cite{chiribella2014dilation}]
	\label{ass:display}   
Every  test arises from a deterministic transformation followed by a measurement on one of the output systems.
\end{ass} 

Formally,  Assumption \ref{ass:display}  is that  every test $\bs{\map T}   =  (\map T_x)_{x\in\set X} $
	of type $\mathrm{A} \to \mathrm{B}$  can be realized as 
	\begin{align}
	\begin{aligned} \Qcircuit @C=1em @R=.7em @!R {  
	& \qw \poloFantasmaCn{\rA}& \gate{\mathcal{T}_{x} } & \qw \poloFantasmaCn{\rB} &\qw     & &=&   & \qw \poloFantasmaCn{\rA}& \multigate{1}{\mathcal{D}} & \qw \poloFantasmaCn{\rB} & \qw   &   \qquad   &\qquad \forall x\in\mathsf{X} \, .
 \\ 
	&&&&&&&  && \pureghost{\mathcal D} & \qw \poloFantasmaCn{\rC} & \measureD{c_x}} && \end{aligned}	\end{align} 
for some system $\mathrm{C}$, some  deterministic transformation $\mathcal{D}  $, and some measurement  $\su c=  (c_x)_{x\in \set X}$.

Our second assumption is that the set of systems  described by the theory includes some ``coin'', that is, some physical system that generates random outcomes. In the familiar setting where probabilities are numbers in the interval $[0,1]$,  the assumption is that there exists   some two-outcome experiment whose  outcome probabilities are both non-zero.    { 
In the general case, an analogue of  non-zero probability is a {\em cancellative  scalar}:  
a scalar $s$  is {\em cancellative}   if the condition $s\,  p  =  s\,  p'$ implies $p=p'$ for every pair of scalars $p$  and $p'$.

\begin{ass}[Coins]\label{ass:Coins}
 There exist  two-outcome experiments whose outcome probabilities are both cancellative.
\end{ass}}
We stress that we do not assume the existence of coins with arbitrary biases: even if the scalars are real numbers, we do not assume that they  are the whole interval $[0,1]$. 

An important property of cancellative scalars is that they cancel out in equations involving arbitrary transformations:  
\begin{lem}\label{lem:cancel}
For a cancellative scalar $s$,  the condition  $s\, \map P  =  s\, \map P'$ implies $\map P= \map P'$, for every pair of processes  $\map P$ and $\map P'$ of the same type.
\end{lem}
\Proof  Given two processes $\map P$ and $\map P'$ of type $\rA \to \rB$, the condition  $s\, \map P  =  s\, \map P'$ implies
\begin{align}
s    
  \begin{aligned}   \Qcircuit @C=1em @R=.7em @!R {  & \multiprepareC{1}{\Gamma} & \qw \poloFantasmaCn{\rA} &  \gate{\map P}  &  \qw \poloFantasmaCn{\rB} & \multimeasureD{1}{E}  \\  
 & \pureghost{\Gamma} & \qw \poloFantasmaCn{\rC} &  \qw &\qw     &\ghost{E}} 
 \end{aligned}  
 \quad =   
  s  
   \begin{aligned} \Qcircuit @C=1em @R=.7em @!R { & \multiprepareC{1}{\Gamma} & \qw \poloFantasmaCn{\rA} &  \gate{\map P'}  &  \qw \poloFantasmaCn{\rB} & \multimeasureD{1}{E}   &\qquad &\qquad & \\ 
	 & \pureghost{\Gamma} & \qw \poloFantasmaCn{\rC} &  \qw &\qw   &  \ghost{E}  &  \qquad  & \qquad &  \qquad } \end{aligned}	
	 &  \begin{array}{c} \forall \rC \in \Sys  \\  \forall \Gamma  \in \St (\rA\otimes \rC)\\
	 \forall  E \in \Eff (\rB\otimes \rC) ~, \end{array}   
		 \end{align}  
 which in turn implies 
\begin{align}
\begin{array}{c} \begin{aligned} \Qcircuit @C=1em @R=.7em @!R {  & \multiprepareC{1}{\Gamma} & \qw \poloFantasmaCn{\rA} &  \gate{\map P}  &  \qw \poloFantasmaCn{\rB} & \multimeasureD{1}{E}  \\  
	 & \pureghost{\Gamma} & \qw \poloFantasmaCn{\rC} &  \qw &\qw     &\ghost{E}} \end{aligned} \quad   =   
   \begin{aligned} \Qcircuit @C=1em @R=.7em @!R { & \multiprepareC{1}{\Gamma} & \qw \poloFantasmaCn{\rA} &  \gate{\map P'}  &  \qw \poloFantasmaCn{\rB} & \multimeasureD{1}{E}   &\qquad &\qquad & \\ 
	 & \pureghost{\Gamma} & \qw \poloFantasmaCn{\rC} &  \qw &\qw   &  \ghost{E}  &  \qquad  & \qquad &  \qquad } \end{aligned}	   \begin{array}{c} \forall \rC \in \Sys  \\  \forall \Gamma  \in \St (\rA\otimes \rC)\\
	 \forall  E \in \Eff (\rB\otimes \rC)  \end{array}   \end{array}
	 \end{align}  
    because $s$ is cancellative. Hence, the processes $\map P$ and $\map P'$ are operationally indistinguishable, in the sense of  Definition \ref{def:opind}. Since we are dealing with a quotient theory, this condition implies $\map P= \map P'$.  \qed

 Our third assumption is that it is  possible to perform randomized tests.  Informally, a randomized test is a test where one tosses a coin and performs a test depending on the outcome of the coin toss.   Here we give the formal definition for the randomization of two tests (the extension to more than two tests is straightforward):  
 \begin{defi} Let $\bs {\map T}$ and  $\bs {\map S}$ be two tests of type $\rA \to \rB$, with outcomes in $\set X$ and $\set Y$, respectively, and let $ (p_0,p_1)$ be a probability distribution allowed by the theory ({\em i.e.} a test of type $\rI\to \rI$).    The randomization of  $\bs {\map T}$ and  $\bs {\map S}$ with probabilities $(p_0,p_1)$ is a test  $\bs {\map R}$ of type $\rA \to \rB$,  with outcomes in the disjoint union $\set Z  =\set X  \sqcup \set Y$, in which all elements of $\set X$ are regarded as distinct from all elements of $\set Y$. The test $\bs {\map R}$ is defined as  
\begin{align}
\map R_z    =  
\left \{     
\begin{array}{ll}
p_0\, \map T_z  \, \qquad  \qquad &   z\in\set X    \\  
&  \\
p_1  \, \map S_z  \qquad \qquad &  z\in\set Y     \, .    
\end{array} 
\right.
\end{align}
\end{defi}
Our third assumption is that all randomizations are valid tests: 

\begin{ass}[Randomizations]\label{ass:random}
The set of tests is closed under randomizations.
\end{ass}

The Randomizations Assumption, together with the possibility of coarse graining,  guarantees the existence of random mixtures, defined as follows:
\begin{defi}
Let $\sigma $ and $\tau$ be two deterministic states of system $\rA$ and let $(p_0,p_1)$ be a probability distribution allowed by the theory.  
The {\em mixture} of   $\sigma$ and $\tau$ with probabilities $(p_0,p_1)$ is the deterministic state $\rho$ defined by \footnote{The fact that $\rho$ is a deterministic state follows from the fact that $\rho$ is the coarse-graining of the randomized test  $\bs \rho  =  (\rho_0,\rho_1)$ with $\rho_0  =  p_0\,    \sigma$ and $\rho_1  =  p_1\,    \tau$. }
\begin{align}
\rho:   =  p_0 \,  \sigma  +  p_1 \, \tau 
\end{align} 
\end{defi}

The notion of mixture introduces a pre-order relation on the set of deterministic states:  
\begin{defi}\label{def:contains}
Let $\rho$ and $\sigma$ be two   deterministic  states of system $\rA$.  We say that $\rho$ {\em contains} $\sigma$, denoted as $\rho  \sqsupseteq \sigma$, if  $\rho  = p_0  \, \sigma +   p_1\, \tau$, where  {  $(p_0,p_1)$ is a probability distribution allowed by the theory,  $p_0$ is a  cancellative scalar, and $\tau$ is a deterministic  state.   }
\end{defi}  

Note that the relation $\sqsupseteq$ is transitive: if $\rho \sqsupseteq \sigma$ and $\sigma \sqsupseteq \tau$, then $  \rho  \sqsupseteq \tau$.

Our final assumption is about the existence of  {complete states}, defined as follows:
\begin{defi}\label{def:complete}
A deterministic state $\rho  \in  \Det\St(\rA)$ is {\em complete} if $\rho  \sqsupseteq \sigma$ for every deterministic state $\sigma \in  \Det\St (\rA)$.  
\end{defi}
The assumption is:
\begin{ass}[Complete States]\label{ass:complete}
For every system, there exists at least one complete state. 
\end{ass}
While Assumptions \ref{ass:display}, \ref{ass:Coins}, and \ref{ass:random} are satisfied by most probabilistic theories, including infinite dimensional ones,   Assumption \ref{ass:complete} is more specific to the finite dimensional setting. 
  It is  satisfied in the standard scenario where  the set of deterministic states is a finite-dimensional convex set.  In this scenario,  the complete states  are exactly the points in the interior of the convex set.   Assumption  \ref{ass:complete} is  also satisfied in some theories where the set of deterministic states is not convex.  For example, it is satisfied in  Spekkens' toy theory \cite{spekkens2007evidence,spekkens2016quasi}, where some convex combinations are forbidden but nevertheless the set of deterministic states  contains a complete state.  However, Assumption  \ref{ass:complete}   is generally not satisfied by classical and quantum theory infinite dimensions.   An infinite-dimensional generalization of the notion of complete state will be discussed in Subsection \ref{subsec:tomoord}. 

\subsection{Local Tomography }  

In this section we review the property of Local Tomography. The scope of this review is to clarify, by contrast, what it means to characterize physical processes  in theories where  Local Tomography  does not hold.  

The Local Tomography  has been formulated in several  forms   and under different names \cite{araki1980characterization,wootters1990local,hardy2001quantum,dariano2006how,barnum2007generalized,barrett2007information,chiribella2010probabilistic,hardy2011reformulating}. A simple formulation is the following \cite{chiribella2012quantum}
\begin{defi}[Local Tomography]
A physical theory satisfies Local Tomography  if the state of  every composite system is uniquely determined by the joint statistics  of  local measurements on the components. 
\end{defi}
Mathematically, Local Tomography  states that,  for every pair of systems $\rA$ and $\rB$, and for every pair of states of the composite system $\rA\otimes \rB$, say $\rho$ and $\rho'$,
 the  condition 
\begin{align}\label{equalityLT}
\begin{aligned} 
\Qcircuit @C=1em @R=.7em @!R {
	 & \multiprepareC{1}{\rho} & \qw \poloFantasmaCn{\rA}   & \measureD{a}  \\ 
	 & \pureghost{\rho} & \qw \poloFantasmaCn{\rB} &  \measureD{b} 
	 }
\end{aligned}    
\quad = 
\begin{aligned} 
\Qcircuit @C=1em @R=.7em @!R {
	 & \multiprepareC{1}{\rho'} & \qw \poloFantasmaCn{\rA}   & \measureD{a}   &\qquad  &\qquad&&\forall a\in\Eff (\rA)  &\qquad&\qquad & \\ 
	 & \pureghost{\rho'} & \qw \poloFantasmaCn{\rB} &  \measureD{b}   &\qquad &\qquad&& \forall b\in\Eff (\rB)  &\qquad&\qquad &
	 }
\end{aligned}    
\end{align}
 implies the equality   
\begin{align}\label{LT2}
\begin{aligned} 
\Qcircuit @C=1em @R=.7em @!R {
	 & \multiprepareC{1}{\rho} & \qw \poloFantasmaCn{\rA}   & \qw \\ 
	 & \pureghost{\rho} & \qw \poloFantasmaCn{\rB} &  \qw 
	 }
\end{aligned}    
\quad = 
\begin{aligned} 
\Qcircuit @C=1em @R=.7em @!R {
	 & \multiprepareC{1}{\rho'} & \qw \poloFantasmaCn{\rA}   & \qw  \\ 
	 & \pureghost{\rho'} & \qw \poloFantasmaCn{\rB} &  \qw ~~~~~~.
	 }
\end{aligned}   
\end{align}

\subsection{Standard process tomography}

Local Tomography  has an important implication for the task of characterizing physical processes: in every locally tomographic theory, the action of a process on its input system determines the action of the process on all possible composite systems. Explicitly, one has the following 
\begin{prop}[Lemma 14 of \cite{chiribella2010probabilistic}]\label{prop:processtomo}
Suppose that the theory satisfies Local Tomography.  Then, for every pair of systems $\rA$ and $\rB$ and for every pair of processes  $\map P$ and $\map P'$ of type $A\to B$, the condition  
 \begin{align}\label{stan}
  \begin{aligned}
 \Qcircuit @C=1em @R=.7em @!R { & \prepareC{\rho} & \qw \poloFantasmaCn{\rA} &  \gate{\map P}  &  \qw \poloFantasmaCn{\rB} & \qw  &\qquad }
\end{aligned} =  
\begin{aligned}
 \Qcircuit @C=1em @R=.7em @!R { & \prepareC{\rho} & \qw \poloFantasmaCn{\rA} &  \gate{\map P'}  &  \qw \poloFantasmaCn{\rB} & \qw  & \qquad &\qquad  &\qquad & \forall \rho\in  \St (\rA)  &\qquad &\qquad &\qquad & } 
\end{aligned}
 \end{align}
implies the equality  
\begin{align}
\begin{aligned} \Qcircuit @C=1em @R=.7em @!R { & \multiprepareC{1}{\Gamma} & \qw \poloFantasmaCn{\rA} &  \gate{\map P}  &  \qw \poloFantasmaCn{\rB} & \qw  &
 \\ 
	 & \pureghost{\Gamma} & \qw \poloFantasmaCn{\rC} &  \qw &\qw   &\qw  &\qquad} \end{aligned}  =  
	 \begin{aligned} \Qcircuit @C=1em @R=.7em @!R { & \multiprepareC{1}{\Gamma} & \qw \poloFantasmaCn{\rA} &  \gate{\map P'}  &  \qw \poloFantasmaCn{\rB} & \qw   &\qquad  &   &\qquad   &\qquad &  \forall \rC\in\Sys &\qquad &  \\ 
	 & \pureghost{\Gamma} & \qw \poloFantasmaCn{\rC} &  \qw &\qw   &  \qw  &\qquad &\qquad &\qquad &  &\qquad \forall \Gamma  \in  \St(\rA\otimes \rC)\, ,  &   \qquad &	 } \end{aligned}	 
	 \end{align} 
	 or equivalently,  the equality 
\begin{align}\label{sameprocess}
  \begin{aligned}
 \Qcircuit @C=1em @R=.7em @!R { & \qw \poloFantasmaCn{\rA} &  \gate{\map P}  &  \qw \poloFantasmaCn{\rB} & \qw  &\qquad }
\end{aligned} =  
\begin{aligned}
 \Qcircuit @C=1em @R=.7em @!R {  &\qquad& \qw \poloFantasmaCn{\rA} &  \gate{\map P'}  &  \qw \poloFantasmaCn{\rB} & \qw   \quad  . } 
\end{aligned}  
 \end{align}
\end{prop} 
When system $A$ is finite dimensional,  Proposition \ref{prop:processtomo} guarantees that  one can characterize  an arbitrary process of type $A\to B$ by applying  it to a fiducial set of input states of system $A$,   and by  characterizing the output states of system $B$ with a fiducial set of measurements. This result  is the conceptual foundation underpinning   standard process tomography \cite{chuang1997prescription,poyatos1997complete}.  

 {   Interestingly, a  converse of Proposition \ref{prop:processtomo} also holds, under the assumption that the  all states of all composite  systems can be probabilistically generated  by acting locally on one part of a fixed  bipartite state. This point is discussed in Appendix \ref{app:protomo}.  
 
}
  
\subsection{The counterexample of  real-vector-space quantum theory}
 When Local Tomography  does not hold, standard process tomography  may not  work. 
  A simple example can be found in real-vector-space quantum theory  \cite{stueckelberg1960quantum}, a variant of standard  quantum theory  that violates Local Tomography  \cite{araki1980characterization,wootters1990local} and deviates from standard quantum theory in a number  of operational tasks \cite{wootters1990local,chiribella2010probabilistic,wootters2012entanglement,hardy2012limited,aleksandrova2013real,chiribella2013quantum,wootters2014rebit,wootters2016optimal}.    The example, introduced in  \cite{chiribella2016quantum}, involves  two processes,  $\map P$ and $\map P'$, acting on a ``rebit'', that is, a two-dimensional quantum system associated to the vector space $\R^2$.  
    Mathematically, the two processes are  defined by the  linear maps   
  \begin{align}
\nonumber  \map P   (M)   &=   \frac 12  \left(  M +  YMY\right)  \\
   \map P'   (M)   &=   \frac 12  \left(  XMX +  ZMZ\right) \, ,
 \end{align}
 where $M$ is a generic $2\times 2$ matrix and $X := \begin{pmatrix}  0  &1  \\  1&  0 \end{pmatrix},Y:= \begin{pmatrix}  0  &-i  \\  i&  0 \end{pmatrix},Z := \begin{pmatrix}  1  &0  \\  0&  -1 \end{pmatrix}$ are the three Pauli matrices.  Now, consider the action of the two processes on a single-rebit, whose possible states are described by density matrices with real entries.  Writing a generic real-valued  density matrix $\rho$ in the Bloch form 
 \begin{align}
 \rho   =   \frac {  I    +    \cos \theta  \,X  +   \sin \theta  \, X }{2}   \, , \qquad \theta \in  [0,2\pi)
 \end{align} 
 we can easily obtain the relation  
 \begin{align}
 \map P(\rho)    =  \frac {I}2    =  \map P'  (\rho) \, , 
 \end{align}
 meaning that the two processes $\map P$ and $\map P'$ act in the same way on every single-rebit input state.  
 
 On the other hand,  the  processes $\map P$ and $\map P'$ are clearly different.  The difference can be detected by applying the processes on a maximally entangled state of {\em two} rebits:  for example, using the notation 
\begin{align}
\nonumber |\Phi^+\>    & :=  \frac {  |0\>\otimes |0\> +  |1\>\otimes |1\> }{\sqrt 2}  \\
\nonumber |\Phi^-\>   &:=  \frac {  |0\>\otimes |0\>   -|1\>\otimes |1\> }{\sqrt 2}  \\
\nonumber |\Psi^+\>   &:=  \frac {  |0\>\otimes |1\> +  |1\>\otimes |0\> }{\sqrt 2}  \\
|\Psi^-\>   &:=  \frac {  |0\>\otimes |1\> -  |1\>\otimes |0\> }{\sqrt 2}   \,.
\end{align}
one has
  \begin{align}\label{bellmix}
\nonumber (\map P \otimes \map I)   (  |\Phi^+\>\<\Phi^+|)   &=  \frac 12  \,      |\Phi^+\>\<\Phi^+| +    |\Psi^-\>\<\Psi^-|  \\
(\map P \otimes \map I)   (  |\Phi^+\>\<\Phi^+|)   &=  \frac 12  \,      |\Psi^+\>\<\Psi^+| +    |\Phi^-\>\<\Phi^-|   \, .
\end{align}
 Note that the output states are not only different, but also {\em orthogonal}, meaning that a joint measurement on the two rebits can tell the two states apart without any error.  The two states in the r.h.s. of Eq. (\ref{bellmix}) were first studied  by Wootters \cite{wootters1990local} as an example of bipartite states that perfectly distinguishable by global measurements,  and  yet completely indistinguishable by local measurements in real-vector-space quantum theory. 

Summarizing, standard process tomography does not work  in real-vector-space quantum theory: in order to unambiguously characterize an unknown process on real-vector space quantum states, it is mandatory to test it on composite systems.

\subsection{
Process tomography without the assumption of Local Tomography}

In the lack of Local Tomography, process tomography requires experiments on composite systems. Now, the key question is: which composites have to be tested?  In principle, the answer could be ``all'': a process with input system $\rA$ may have to be tested on every  composite system $\rA \otimes \rC$ for every possible auxiliary system $\rC$. 
Loosely speaking, the only way to completely characterize a process would be to make experiments on ``the whole universe''.

This   situation does not arise  in real-vector-space quantum theory: there, a complete process tomography can be achieved by preparing a {\em single} input state of a {\em single} composite system.  Specifically,  process tomography can be achieved by preparing two identical copies of the input system in a maximally entangled state  and by letting the process act on one copy \footnote{We do not provide a proof here because later in the paper we will give a general proof valid for all  theories obeying the Purification Principle (see subsection \ref{subsec:purification} for the formal definition).  Since real-vector-space quantum theory  satisfies the Purification Principle, the general proof applies.   }. 
 In other words, real-vector-space quantum theory satisfies the principle of Dynamically Faithful States (Principle \ref{prin:dynfaith1} in the introduction).  

In the rest of the paper we will investigate the conditions that guarantee  the validity of  Dynamically Faithful States  in general physical theories.

\section{Results}

 \subsection{Four levels of process tomography}

The goal of  tomography is to characterize the action of an unknown process on {\em all} possible inputs.    
Nevertheless, it is also useful to first  consider intermediate tasks where the goal is to characterize the process on a subset of inputs. 
In this Section we define three such tasks, listed in order of increasing strength.    All together, these three tasks and the  task of full process tomography define four levels of characterization of physical processes.

\subsubsection{Equality on a source}
Consider the situation where a  source prepares system $\rA$ in a non-deterministic fashion. Such a source can be described by a preparation test $\bs \rho  =  (\rho_x)_{x\in\set X}$, where the preparation of the state $\rho_x$  is heralded by the outcome $x$.     
The states in the  source are then  used as inputs: 
\begin{defi}
We say that two processes $\map P$ and $\map P'$, of type $\rA\to \rB$, are {\em equal on the source}  $(\rho_x)_{x\in\set X}$ if one has 
\begin{align}
  \begin{aligned}
 \Qcircuit @C=1em @R=.7em @!R { & \prepareC{\rho_x} & \qw \poloFantasmaCn{\rA} &  \gate{\map P}  &  \qw \poloFantasmaCn{\rB} & \qw }
\end{aligned}  \quad=  
\begin{aligned}
 \Qcircuit @C=1em @R=.7em @!R { & \prepareC{\rho_x} & \qw \poloFantasmaCn{\rA} &  \gate{\map P'}  &  \qw \poloFantasmaCn{\rB} & \qw  & \qquad  &\qquad &  \forall x\in\set X ~. &\qquad &\qquad &\qquad &   \qquad &\qquad} 
\end{aligned}
 \end{align}
When this is the case, we write $\map P  \sim_{\bs \rho}  \map P'$.  
\end{defi}  
It is easy to see that $\sim_{\bs \rho}$ is an equivalence relation.  
The equivalence relation $\sim_{\bs \rho}$  defines a weak notion of process tomography, where the goal is just to identify the action  of  processes  on the fixed  set of states $\{  \rho_x  \, ,  x\in\set X\}$.  

\subsubsection{Equality upon input of a state}
 
We now consider a  stronger notion of process tomography.  Instead of characterizing a process on a single source, one can try to  characterize the process on {\em all}  sources with  the same average state.     The {\em average state of a source} $\bs \rho  =  (\rho_x)_{x\in\set X}$ is the deterministic state 
\begin{align}
\rho   :  =  \sum_{x\in\set X} \, \rho_x   \, ,
\end{align}  
obtained by coarse-graining over all possible outcomes.

  \begin{defi}
We say that two processes $\map P$ and $\map P'$, of type $\rA\to \rB$, are {\em equal upon input of}  $ \rho$ if they are equal on every source  $\bs \rho = (\rho_x)_{x\in\set X}$ with average state equal to $\rho$.  
When this is the case, we write $\map P  =_{\rho}  \map P'$.  
\end{defi} 
The condition $\map P  =_{\rho}  \map P'$ is equivalent to the condition  
\begin{align}
  \begin{aligned}
 \Qcircuit @C=1em @R=.7em @!R { & \prepareC{\sigma} & \qw \poloFantasmaCn{\rA} &  \gate{\map P}  &  \qw \poloFantasmaCn{\rB} & \qw }
\end{aligned}  \quad=  
\begin{aligned}
 \Qcircuit @C=1em @R=.7em @!R { & \prepareC{\sigma} & \qw \poloFantasmaCn{\rA} &  \gate{\map P'}  &  \qw \poloFantasmaCn{\rB} & \qw  & \qquad  &\qquad &  \qquad &  \forall \sigma:  ~ \sigma   \sqsubseteq  \rho  ~, } 
\end{aligned} 
 \end{align}
 where the notation $\sigma \sqsubseteq \rho$ means that $\sigma$ is {\em contained} in $\rho$, that is, that there exists a source $\bs \rho  =  (\rho_0,\rho_1)$ such that  $\rho_0+\rho_1 =  \rho$ and  $\rho_0   =  p_0  \, \sigma$ for some cancellative scalar $p_0$ \footnote{Note that this definition  is slightly more general than Definition \ref{def:contains}:  here do not require the state $\sigma$ to be deterministic. }.

  In quantum information, the notion of equality upon input of $\rho$ was introduced in \cite{nielsen2000quantum}. Its extension to general probabilistic theories was discussed in \cite{chiribella2010probabilistic,chiribella2011informational,dariano2017quantum}.    
  
  It is easy to check that equality upon input of $\rho$ is an equivalence relation. 
Note that, by definition,  identifying a process upon input of $\rho$ is more demanding than just identifying it   on a specific source $(\rho_x)_{x\in\set X}$ with average state $\rho$.

 \subsubsection{Equality on the extensions of a state}
 All notions of process tomography considered so far focussed on the action of a process  on its input system alone.  To identify the process, however, one also needs to characterize its  local action on composite systems. 
      \begin{defi}
 Let $\rho  \in   \Det\St (\rA)$ be a deterministic state of system $\rA$.  An {\em extension} of    $\rho$ on system $\rE$  is a deterministic  state $ \Gamma  \in  \Det\St (\rA\otimes \rE) $ satisfying the relation
 	\begin{align}
	\begin{aligned} \Qcircuit @C=1em @R=.7em @!R {  
	  &   \multiprepareC{1}{\Gamma} & \qw \poloFantasmaCn{\rA} & \qw       & &=&   & \prepareC{\rho } & \qw \poloFantasmaCn{\rA} &\qw 
 \\ 
	& \pureghost{\Gamma} & \qw \poloFantasmaCn{\rE} & \measureD{e}} &  & &  &&  &   \end{aligned}	\end{align} 
for  some deterministic effect  $ e \in  \Det\Eff (\rE)$.    We denote by $\Ext (\rho,\rE)$ the set of all extensions of $\rho$ on system $\rE$.
 \end{defi}    
 
Our  third level  of process tomography is  to identify the action of a process on all the extensions of a given state.   
\begin{defi}
We say that two processes  $\map P$ and $\map P'$ are {\em equal on the extensions of $\rho$}
if one has 
\begin{align}\label{equalext}
\begin{aligned} \Qcircuit @C=1em @R=.7em @!R { & \multiprepareC{1}{\Gamma} & \qw \poloFantasmaCn{\rA} &  \gate{\map P}  &  \qw \poloFantasmaCn{\rB} & \qw 
 \\ 
	 & \pureghost{\Gamma} & \qw \poloFantasmaCn{\rE} &  \qw &\qw   &\qw} \end{aligned} \quad   =  
	 \begin{aligned} \Qcircuit @C=1em @R=.7em @!R { & \multiprepareC{1}{\Gamma} & \qw \poloFantasmaCn{\rA} &  \gate{\map P'}  &  \qw \poloFantasmaCn{\rB} & \qw  
 \\ 
	 & \pureghost{\Gamma} & \qw \poloFantasmaCn{\rE} &  \qw &\qw   &  \qw} \end{aligned}	
	 \end{align}
 for every possible system $\rE$ and for every possible extension $\Gamma  \in \Ext (\rho , \rE)$.  In this case, we write   $\map P \equiv_\rho  \map P'$.
\end{defi}


In general, equality on all the extensions of $\rho$  implies equality upon input of $\rho$. This is because all sources with average state $\rho$ can be viewed as extensions of $\rho$  involving  an environment serving as a ``display''. Explicitly, we have the following:
\begin{lem}[A1]\label{lem:intermediate}
Let $\rho  \in  \Det\St(\rA)$ be a deterministic state of system $\rA$, and let $\bs \rho  =  (\rho_x)_{x\in\set X}$ be an arbitrary source with average state $\rho$.  Then, there exists an extension  $\Gamma \in \Ext (\rho, \rE)$, for suitable system $\rE$, such that, for every pair of processes $\map P$ and $\map P'$ of type $\rA\to \rB$, with arbitrary system $\rB$,   the condition  
\begin{align}
\begin{aligned} \Qcircuit @C=1em @R=.7em @!R { & \multiprepareC{1}{\Gamma} & \qw \poloFantasmaCn{\rA} &  \gate{\map P}  &  \qw \poloFantasmaCn{\rB} & \qw 
 \\ 
	 & \pureghost{\Gamma} & \qw \poloFantasmaCn{\rE} &  \qw &\qw   &\qw} \end{aligned} \quad   =  
	 \begin{aligned} \Qcircuit @C=1em @R=.7em @!R { & \multiprepareC{1}{\Gamma} & \qw \poloFantasmaCn{\rA} &  \gate{\map P'}  &  \qw \poloFantasmaCn{\rB} & \qw  
 \\ 
	 & \pureghost{\Gamma} & \qw \poloFantasmaCn{\rE} &  \qw &\qw   &  \qw} \end{aligned}	
	 \end{align}
  implies the condition  
  \begin{align}
  \begin{aligned}
 \Qcircuit @C=1em @R=.7em @!R { & \prepareC{\rho_x} & \qw \poloFantasmaCn{\rA} &  \gate{\map P}  &  \qw \poloFantasmaCn{\rB} & \qw }
\end{aligned}  \quad=  
\begin{aligned}
 \Qcircuit @C=1em @R=.7em @!R { & \prepareC{\rho_x} & \qw \poloFantasmaCn{\rA} &  \gate{\map P'}  &  \qw \poloFantasmaCn{\rB} & \qw  & \qquad  &\qquad &  \forall x\in\set X ~. &\qquad &\qquad &\qquad &   \qquad &\qquad} 
\end{aligned}
 \end{align}
\end{lem}
The proof is provided in Appendix \ref{app:LT}. An immediate consequence is that equality on all the extensions of $\rho$  implies equality upon input of $\rho$: 
\begin{cor}[A1]\label{cor:deterministic}
Let $\rho  \in  \Det\St(\rA)$ be a deterministic state of system $\rA$.  Then,  for every pair of processes $\map P$ and $\map P'$ of type $\rA\to \rB$, with arbitrary system $\rB$, one has the implication 
\begin{align}
\map P     \equiv_\rho   \map P' 
  \qquad \Longrightarrow \qquad \map P  =_\rho   \map P'  \, . 
\end{align}
\end{cor}

In Appendix \ref{app:LT}  we also   show that  the relations $\equiv_\rho$ and $=_\rho$ coincide in the special case of theories satisfying  Local Tomography  and  Causality, the principle that   the outcome probabilities of present experiments are independent on the choice of future experiments   \cite{chiribella2010probabilistic,coecke2010causal,chiribella2011informational,chiribella2012quantum,coecke2013causal,chiribella2014dilation,chiribella2016quantum,dariano2017quantum,coecke2018picturing}.   


 \subsubsection{Equality on all states}
 
The ultimate goal of process tomography is to identify the action of an unknown process on all possible states.  Using the results collected so far, the condition for perfect identification can be expressed as follows: 
 \begin{lem}[A1]\label{lem:equalonextensionsofdeterministic}
 Let $\map P$ and $\map P'$ be two processes of type $\rA \to \rB$.  Then, one has the condition
 \begin{align}
 \map P  = \map P'  \qquad \Longleftrightarrow \qquad  \map P  \equiv_\rho  \map P'   \, , \quad \forall \rho\in\Det\St(\rA)  \, .
 \end{align}
 \end{lem}

The proof is provided in Appendix \ref{app:extensiondet}. Summarizing, two processes of type $\rA\to \rB$ are equal if and only if they coincide on all the extensions of all deterministic states of system $\rA$.   In Subsection \ref{subsec:mainresult} we will show that,  in fact,  the extensions of  a single state are enough.


  \subsection{Tomographic ordering and dynamically faithful states}\label{subsec:tomoord}  


A natural notion in the context of process tomography is the notion of {\em tomographic ordering}: 
\begin{defi}
Let $\Phi  \in  \St  (\rA\otimes \rR)$ and $\Psi \in \St (\rA \otimes \rE)$ be two (possibly non-deterministic)  states.
We say that $\Phi$ is {\em tomographically more powerful than} $\Psi$   {\em  for processes of type $\rA\to \rB$}, if, for every pair of processes $\map P$ and $\map P'$ of type $\rA\to \rB$,   the condition 
\begin{align}\label{dinfaith}
\begin{aligned} \Qcircuit @C=1em @R=.7em @!R { & \multiprepareC{1}{\Phi} & \qw \poloFantasmaCn{\rA} &  \gate{\map P}  &  \qw \poloFantasmaCn{\rB} & \qw 
 \\ 
	 & \pureghost{\Phi} & \qw \poloFantasmaCn{\rR} &  \qw &\qw   &\qw} \end{aligned} \quad   =  
	 \begin{aligned} \Qcircuit @C=1em @R=.7em @!R { & \multiprepareC{1}{\Phi} & \qw \poloFantasmaCn{\rA} &  \gate{\map P'}  &  \qw \poloFantasmaCn{\rB} & \qw  
 \\ 
	 & \pureghost{\Phi} & \qw \poloFantasmaCn{\rR} &  \qw &\qw   &  \qw} \end{aligned}	
	 \end{align}
implies the condition
\begin{align}
\begin{aligned} \Qcircuit @C=1em @R=.7em @!R { & \multiprepareC{1}{\Psi} & \qw \poloFantasmaCn{\rA} &  \gate{\map P}  &  \qw \poloFantasmaCn{\rB} & \qw 
 \\ 
	 & \pureghost{\Psi} & \qw \poloFantasmaCn{\rE} &  \qw &\qw   &\qw} \end{aligned} \quad   =  
	 \begin{aligned} \Qcircuit @C=1em @R=.7em @!R { & \multiprepareC{1}{\Psi} & \qw \poloFantasmaCn{\rA} &  \gate{\map P'}  &  \qw \poloFantasmaCn{\rB} & \qw  
 \\ 
	 & \pureghost{\Psi} & \qw \poloFantasmaCn{\rE} &  \qw &\qw   &  \qw& \quad  .} \end{aligned}	
	 \end{align}
 When this is the case, we write $\Phi  \succeq_{\rA\to \rB}  \Psi$.  
\end{defi}

 
 


In general, the relation of tomographic ordering (for a fixed systems $\rA$ and $\rB$) may or may not have a maximum. When it does, this maximum is called a {\em  dynamically faithful state}\footnote{Originally \cite{dariano2006missing,dariano2006how,dariano2007operational,maurobook}, the term ``dynamically faithful state'' was used  to refer to states that determine  the action of an unknown process on its input system, up to probabilistic rescalings, {\em i.e.}  states $\Phi$ for which the condition  (\ref{dinfaith}) implies $\map P  \rho  \propto   \map P'\rho$ for every state $\rho\in\St (\rA)$.   
 Later \cite{chiribella2010probabilistic,chiribella2011informational,chiribella2016quantum,dariano2017quantum},   the term was used to denote  states that identify the action of the process on arbitrary composite systems, as per Definition \ref{def:dynfaith}.} 
:  

\begin{defi}[Dynamically faithful state \cite{dariano2006missing,dariano2006how,dariano2007operational,maurobook,chiribella2010probabilistic,chiribella2011informational,chiribella2016quantum,dariano2017quantum}]\label{def:dynfaith}
A (possibly non-deterministic) state $\Phi \in  \St (\rA \otimes \rR)$ is {\em dynamically faithful for processes of type $\rA\to \rB$} if, for every pair of processes $\map P$ and $\map P'$ of type $\rA\to \rB$,  the condition  
\begin{align}
\begin{aligned} \Qcircuit @C=1em @R=.7em @!R { & \multiprepareC{1}{\Phi} & \qw \poloFantasmaCn{\rA} &  \gate{\map P}  &  \qw \poloFantasmaCn{\rB} & \qw 
 \\ 
	 & \pureghost{\Phi} & \qw \poloFantasmaCn{\rR} &  \qw &\qw   &\qw} \end{aligned} \quad   =  
	 \begin{aligned} \Qcircuit @C=1em @R=.7em @!R { & \multiprepareC{1}{\Phi} & \qw \poloFantasmaCn{\rA} &  \gate{\map P'}  &  \qw \poloFantasmaCn{\rB} & \qw  
 \\ 
	 & \pureghost{\Phi} & \qw \poloFantasmaCn{\rR} &  \qw &\qw   &  \qw} \end{aligned}	
	 \end{align}
	 implies 
	 \begin{align}
\begin{aligned} \Qcircuit @C=1em @R=.7em @!R {  & \qw \poloFantasmaCn{\rA} &  \gate{\map P}  &  \qw \poloFantasmaCn{\rB} & \qw }
  \end{aligned} \quad   =  
	 \begin{aligned} \Qcircuit @C=1em @R=.7em @!R { &  \qw \poloFantasmaCn{\rA} &  \gate{\map P'}  &  \qw \poloFantasmaCn{\rB} & \qw \quad .} \end{aligned}	
	 \end{align}
\end{defi}

In every theory satisfying the Displays Assumption \ref{ass:display}, dynamically  faithful states  can be taken to be deterministic without loss of generality:  
\begin{lem}[A1]\label{lem:faithdet}
For a generic pair of systems $\rA$ and $\rB$, if there exists a dynamically faithful state for processes of type $\rA \to \rB$, then  there exists  a {\em deterministic} dynamically faithful state for processes of type  $\rA\to \rB$.
\end{lem}
\Proof
 Let $\Phi  \in \St  (\rA \otimes \rR)$ be a (possibly non-deterministic)  dynamically  faithful state for processes of type  $\rA\to \rB$, and let $ {\bs  {\rho}}  =  (\rho_x)_{x\in\set X}$ be a source that produces the state $\Phi$, say $\rho_{x_0}  = \Phi$ for some outcome $x_0\in \set X$.  Since $\Phi$ is dynamically faithful,     equality  of two processes on the source $\bs \rho$ implies equality of the processes themselves:  for every pair of processes $\map P$ and $\map P'$ of type $\rA\to \rB$, the condition  $(\map P\otimes \map I_\rR)  \sim_{\bs \rho}  (\map P'  \otimes \map I_{\rR})  $  implies the condition $ \map P  =\map P'$.   
 Then, Lemma \ref{lem:intermediate}   shows that there exists a deterministic state $\Gamma  \in \Det\St (\rA \otimes \rR \otimes  \rE)$, for suitable system $\rE$, such that  the condition $(\map   P\otimes \map I_{\rR} \otimes \map I_{\rE}  ) \Gamma  = (\map   P'\otimes \map I_{\rR} \otimes \map I_{\rE}  ) \Gamma $ implies $ (\map P\otimes \map I_\rE)  \sim_{\bs \rho}  (\map P'  \otimes \map I_{\rE}) $ (and therefore $\map P =  \map P'$, in the present case).    
  Hence, the deterministic state $\Gamma$ is dynamically faithful for processes of type $\rA\to \rB$. \qed


\subsection{Relation between tomographic ordering and containment}
There is an important connection between the tomographic ordering and the containment relation defined earlier in the paper. 
Recall the definition:   a deterministic state  $\rho$ contains  another deterministic  state $\sigma$, denoted as $\rho  \sqsupseteq \sigma$, if  $\rho$ is a non-trivial mixture of $\sigma$ and  some other  state $\tau$, namely $\rho  =  p_0 \, \sigma  +  p_1\,  \tau$ where $(p_0,p_1)$ are probabilities  { and  $p_0$ is cancellative.  }
Here we show that the containment relation orders the deterministic states according to the tomographic power of their extensions.  

\begin{lem}[A1,A3]\label{lem:containtomo}
Let $\rho$ and $\sigma$ be two deterministic states of system $\rA$.  If $\rho$ contains $\sigma$, then for every extension of $\sigma$, say $\Sigma  \in \Ext(\sigma,\rE)$ there exists an extension of $\rho$, say $\Gamma  \in \Ext(\rho,\rF)$, such that 
\begin{align}
\Gamma  \succeq_{\rA\to \rB}   \Sigma  
\end{align} 
for every system $\rB$.
\end{lem}
The proof is provided in Appendix \ref{app:containtomo}.

\medskip 

{ A  simple  consequence of Lemma \ref{lem:containtomo} is the following 
\begin{cor}[A1,A3]\label{cor:containtomo}
Let $\rho$ and $\sigma$ be two deterministic states of system $\rA$.  If $\rho$ contains $\sigma$, then, for every system $\rB$ and for every pair of processes $\map P$ and $\map P'$ of type $\rA\to \rB$,  the condition   $\map P \equiv_\rho   \map P'$ implies the condition  $\map P   \equiv_\sigma \map P'$,    
\end{cor}
The case of complete states is especially important here. In this case, we have the following 
\begin{cor}[A1,A3]\label{cor:complete}
If $\omega  \in  \Det\St (\rA)$ is a complete state, then, for every system $\rB$ and every pair of processes $\map P$ and $\map P'$ of type $\rA\to \rB$,  the condition $\map P\equiv_\omega  \map P'$ implies  the equality $\map P  =  \map P'$. 
\end{cor} 
\Proof Immediate from  Corollary \ref{cor:containtomo}, Lemma \ref{lem:equalonextensionsofdeterministic}, and the definition of complete state.\qed   

This result plays a crucial role in the rest of the paper.  In passing, we note that it also suggests an infinite-dimensional generalization of the notion of complete state \footnote{We say that a state  $\omega  \in \Det\St (\rA)$ is {\em tomographically complete} if the relation $\map P \equiv_\omega  \map P'$ implies the equality $\map P=  \map P'$ for every pair of processes of type $\rA\to \rB$ with arbitrary output system $\rB$.   Corollary \ref{cor:complete} shows that every complete state is tomographically complete. In infinite dimensional settings, complete states in the sense of Definition \ref{def:complete} may not exist, and one could regard tomographically complete states as the appropriate notion.   Notably, all the results in this paper hold if the notion of complete state is replaced by the more general notion of tomographically complete state.    }.
  }

\subsection{Necessary and sufficient  condition for Dynamically Faithful States}\label{subsec:mainresult}

{ We  are now ready to provide the core technical result of the paper: 

\begin{theo}[A1,A2,A3,A4]\label{theo:math}
For a generic pair of systems $\rA$ and $\rB$,   the following are equivalent:  
\begin{enumerate}
\item there exists a dynamically faithful state for processes of type $\rA \to \rB$
\item there exists a  complete state $\omega \in \Det\St(\rA)$ such that the set  of all the extensions of $\omega$  has a maximum with respect to the tomographic ordering  $\succeq_{\rA\to \rB}$. 
\end{enumerate}
\end{theo}

The proof is provided in Appendix \ref{app:main}.} The important point of the above theorem is that one can focus the attention on the complete states.  In summary, the existence of dynamically faithful states  boils down to the existence of a   {\em maximally powerful extension} of a  complete state.  



\subsection{Physical conditions for Dynamically Faithful States}

We have seen that Dynamically Faithful States  holds  if and only if every complete state has a maximally powerful extension. Still, this is just a mathematical requirement. Can we find  physical reasons for this requirement to be satisfied?    
 Here we give three physically motivated sufficient conditions. 

\subsubsection{Universal Extension}\label{subsubsec:uniext}
Heuristically, it is natural to imagine that  the ``maximally powerful extension'' in Theorem \ref{theo:math}  determines  all the other  ways in which the system can be correlated to an environment. Going further, we can imagine that all the other correlations can be physically generated from the ``maximally powerful extension'' using some (possibly non-deterministic) process.  The  idea motivates the notion of universal extension, from which all the extensions of a given state can be generated: 
 \begin{defi}[Universal extension]\label{def:uniext}
 A  state $\rho  \in  \St (\rA)$  has a  {\em universal extension}  if there exists a {\em reference system} $\rR$ and a state $\Psi  \in\St(\rA\otimes \rR)$   with the property that every other extension of $\rho$ can be probabilistically generated  from $\Psi$ by applying  a transformation on   $\rR$.   Explicitly:  for every system $\rE$, every extension  $\Gamma \in  \Ext  (\rho, \rE )$ can be obtained  as 
 \begin{align}\label{universal}
p\,  \begin{aligned} \Qcircuit @C=1em @R=.7em @!R {
	 & \pureghost{\Gamma} & \qw \poloFantasmaCn{\rA} &  \qw  \\
	 & \multiprepareC{-1}{\Gamma} & \qw \poloFantasmaCn{\rE}   &\qw  } 
\end{aligned}     
	 \quad    =  \begin{aligned} \Qcircuit @C=1em @R=.7em @!R {
	 & \pureghost{\Psi} & \qw \poloFantasmaCn{\rA} &  \qw  &\qw & \qw   \\
	 & \multiprepareC{-1}{\Psi} & \qw \poloFantasmaCn{\rR}   &\gate{ \map T}  &    \qw \poloFantasmaCn{\rE}  & \qw   } 
\end{aligned}   \, ,
	 \end{align}
where  $p\in\Transf(\rI\to \rI)$ is a cancellative scalar  and $\map T$ is a transformation of type $\rR\to \rE$. 
 \end{defi}     

Using this definition, we can formulate a  requirement for general physical theories: 
\begin{prin}[Universal Extension]\label{prin:uniext}
Every  physical system has a complete state with a universal extension.  
\end{prin}
Note that one could also consider the stronger requirement that {\em every} state have a universal extension.  The reasons why we did not phrase Principle \ref{prin:uniext}  in this way are  (1) universal extensions of complete states are already sufficient for process tomography,  and  (2) as we will see later in the paper, the possibility of conclusive teleportation guarantees the existence of a universal extension of a a complete state, while in general it may not guarantee the existence of universal extensions of arbitrary states.

We now show that Universal Extension guarantees Dynamically Faithful States.   The first step is the following lemma:  
\begin{lem}\label{lem:equalityonallext}
 Suppose that a  deterministic state $\rho  \in \Det\St (\rA)$ has a universal extension $\Psi  \in  \Det\St(\rA\otimes \rR)$.     Then, for every system $\rB$ and for every pair of processes  of type $\rA\to \rB$, say $\map P $ and $\map P'$, the condition  
 \begin{align}\label{uno}
\begin{aligned} \Qcircuit @C=1em @R=.7em @!R { & \multiprepareC{1}{\Psi} & \qw \poloFantasmaCn{\rA} &  \gate{\map P}  &  \qw \poloFantasmaCn{\rB} & \qw 
 \\ 
	 & \pureghost{\Psi} & \qw \poloFantasmaCn{\rR} &  \qw &\qw   &\qw} \end{aligned} \quad   =  
	 \begin{aligned} \Qcircuit @C=1em @R=.7em @!R { & \multiprepareC{1}{\Psi} & \qw \poloFantasmaCn{\rA} &  \gate{\map P'}  &  \qw \poloFantasmaCn{\rB} & \qw  
 \\ 
	 & \pureghost{\Psi} & \qw \poloFantasmaCn{\rR} &  \qw &\qw   &  \qw} \end{aligned}	
	 \end{align}
 implies the condition   
 \begin{align}
 \map P  \equiv_\rho  \map P'  \, .
 \end{align}
\end{lem}     

\Proof Let $\Gamma \in \Ext  (  \rho,  \rE)$ be an arbitrary extension of $\rho$, and let $ \map T \in \Transf(\rR\to \rE)$ be the transformation in Eq.~(\ref{universal}).    By applying the transformation $\map T$ on both sides of  Eq. (\ref{uno}), we obtain   the relation
\begin{align}
 \begin{aligned} \Qcircuit @C=1em @R=.7em @!R { & \multiprepareC{1}{\Psi} & \qw \poloFantasmaCn{\rA} &  \gate{\map P}  &  \qw \poloFantasmaCn{\rB} & \qw 
 \\ 
	 & \pureghost{\Psi} & \qw \poloFantasmaCn{\rR} &  \gate{\map T}&\qw \poloFantasmaCn{\rE}   &\qw} \end{aligned} \quad   =    
	 \begin{aligned} \Qcircuit @C=1em @R=.7em @!R { & \multiprepareC{1}{\Psi} & \qw \poloFantasmaCn{\rA} &  \gate{\map P'}  &  \qw \poloFantasmaCn{\rB} & \qw  
 \\ 
	 & \pureghost{\Psi} & \qw \poloFantasmaCn{\rR} &  \gate{\map T} &\qw  \poloFantasmaCn{\rE}  &  \qw} \end{aligned}	~.
	 \end{align}
Inserting Eq.  (\ref{universal}) into both sides of the equality, we then obtain

\begin{align}
p\, \begin{aligned} \Qcircuit @C=1em @R=.7em @!R { & \multiprepareC{1}{\Gamma} & \qw \poloFantasmaCn{\rA} &  \gate{\map P}  &  \qw \poloFantasmaCn{\rB} & \qw 
 \\ 
	 & \pureghost{\Gamma} & \qw \poloFantasmaCn{\rE} &  \qw &\qw   &\qw} \end{aligned} \quad   =  
p\, 	 \begin{aligned} \Qcircuit @C=1em @R=.7em @!R { & \multiprepareC{1}{\Gamma} & \qw \poloFantasmaCn{\rA} &  \gate{\map P'}  &  \qw \poloFantasmaCn{\rB} & \qw  
 \\ 
	 & \pureghost{\Gamma} & \qw \poloFantasmaCn{\rE} &  \qw &\qw   &  \qw} \end{aligned}	~ ,
	 \end{align}
	 and, since $p$ is cancellative,  
	 \begin{align}
 \begin{aligned} \Qcircuit @C=1em @R=.7em @!R { & \multiprepareC{1}{\Gamma} & \qw \poloFantasmaCn{\rA} &  \gate{\map P}  &  \qw \poloFantasmaCn{\rB} & \qw 
 \\ 
	 & \pureghost{\Gamma} & \qw \poloFantasmaCn{\rE} &  \qw &\qw   &\qw} \end{aligned} \quad   =  
 \begin{aligned} \Qcircuit @C=1em @R=.7em @!R { & \multiprepareC{1}{\Gamma} & \qw \poloFantasmaCn{\rA} &  \gate{\map P'}  &  \qw \poloFantasmaCn{\rB} & \qw  
 \\ 
	 & \pureghost{\Gamma} & \qw \poloFantasmaCn{\rE} &  \qw &\qw   &  \qw} \end{aligned}	
	 \end{align}
	 (cf. Lemma \ref{lem:cancel}).
	 Since $\Gamma$ is an arbitrary extension of $\rho$, we conclude that the equivalence $\map P  \equiv_\rho  \map P$ holds.  
 \qed

Applying the above lemma to a complete state, we obtain the following 
\begin{theo}[A1,A3]\label{theo:uniext}
Universal Extension implies Dynamically Faithful States.   
 \end{theo}  

\Proof   Universal Extension guarantees that there exists a complete state $\omega  \in \Det\St (\rA)$ with a universal extension $\Phi\in \Det\St (\rA\otimes \rR)$, for some auxiliary system $\rR$.    Since $\Phi$ is a universal extension,  for every system $\rB$ and every pair of processes $\map P$ and $\map P'$ of type $\rA \to \rB$,  the equality \begin{align}
 (\map P \otimes \map I_\rR)  \Phi    =   (\map P' \otimes \map I_\rR)  \Phi 
 \end{align}
 implies the condition   
 \begin{align}
 \map P  \equiv_\omega  \map P'  
 \end{align} 
(cf. Lemma \ref{lem:equalityonallext}).    Now, since $\omega$ is complete, the  equality on the extensions of $\omega$ implies  $\map P = \map P'$ (Corollary  \ref{cor:complete}). 
  Summarizing, we obtained that the condition  $(\map P \otimes \map I_\rR)  \Phi    =   (\map P' \otimes \map I_\rR)  \Phi $ implies $\map P=  \map P'$. Hence, the state $\Phi$ is dynamically faithful for processes of type $\rA\to \rB$. Since this construction works for arbitrary $\rA$ and  $\rB$, we conclude that the theory satisfies   Dynamically Faithful States.  \qed

Notably, Universal Extension implies the existence of a state $\Phi \in  \Det\St(\rA\otimes \rB)$ which is dynamically faithful for processes of type $\rA \to \rB$,  
 with {\em arbitrary} output system $\rB$. In other words, the choice  dynamically faithful state depends only on the input  of the tested process and not on its output.

 \subsubsection{Conclusive Teleportation}
 
 Quantum teleportation~\cite{bennett1993teleporting} is a protocol for transferring the state of a quantum system from a sender to a receiver using only classical communication and   previously established quantum correlations. The most basic version of quantum teleportation is probabilistic: the sender performs a joint measurement  on the system to be teleported and  on  part of a bipartite state, shared with the receiver. If the measurement yields a specific outcome, then  the state of the system is transferred to the receiver without any alteration.  Probabilistic teleportation is often called {\em conclusive}.   The analogue of conclusive  teleportation in general  theories was studied by Abramsky and Coecke in \cite{abramsky2004,coecke2006kindergarten} in the context of categorical quantum mechanics,  by Barnum, Barrett, Leifer, and Wilce  \cite{barnum2007generalized} in the context of general probabilistic  theories, and by  Chiribella, D'Ariano, and Perinotti \cite{chiribella2010probabilistic,dariano2017quantum} in the context of theories satisfying Purification.  

In a general physical theory, conclusive teleportation  means that, for every system $\rA$ there exists a system $\rR$, a deterministic state $\Phi\in\Det\St(\rA\otimes \rR) $, an effect  $E  \in  \Eff (\rR\otimes \rA)$, and a cancellative scalar $p_\rA$ such that  
 \begin{align}\label{teleportation}
\begin{aligned} \Qcircuit @C=1em @R=.7em @!R {
	 & \pureghost{\Phi} & \qw \poloFantasmaCn{\rA} &  \qw    \\
	 & \multiprepareC{-1}{\Phi} & \qw \poloFantasmaCn{\rR}   &\multimeasureD{1}{E}  \\
	  & & \qw \poloFantasmaCn{\rA}   &\ghost{E}      } 
\end{aligned}    
\quad = \quad p_\rA   \quad
\begin{aligned} \Qcircuit @C=1em @R=.7em @!R {
 & \qw \poloFantasmaCn{\rA} &\gate{\map I_\rA }  & \qw \poloFantasmaCn{\rA}  & \qw }   
\end{aligned}	     ~.
\end{align}
The existence of conclusive teleportation is a valuable operational feature,    and may be put forward as a desideratum for general physical theories:
\begin{prin}[Conclusive Teleportation]
For every physical system $\rA$, there exists a conclusive teleportation protocol as in Eq. (\ref{teleportation}).  
\end{prin}
Modulo a few technical differences, Conclusive Teleportation coincides with the main axiom  for compact closed categories in categorical quantum mechanics \cite{abramsky2004,coecke2006kindergarten,abramsky2008,coecke2010quantum,coecke2018picturing}.    In this context, it is known that the teleportation state $\Phi$ induces a one-to-one correspondence between processes and bipartite states. In other words,  $\Phi$ is   dynamically faithful for processes of type $\rA \to \rB$, with arbitrary $\rB$.    Here we show a stronger result, namely that the state $\Phi$ is the universal extension of a complete state:  

\begin{prop}[A4]\label{prop:tele}
Let $\Phi \in \Det\St (\rA\otimes \rR)$ be a teleportation state for system $\rA$. Then, $\Phi$ is the universal extension of a complete state.    
\end{prop} 

The proof is provided in Appendix \ref{app:tele}, where we also observe that, in general, a theory with conclusive teleportation may not have universal extensions for arbitrary states.

A simple consequence of Proposition  \ref{prop:tele} is the following: 
\begin{cor}[A4]
Conclusive Teleportation implies  Universal Extension. 
\end{cor}

Summarizing, we proved the chain of implications: 
\begin{align*}
\nonumber {\rm Conclusive~Teleportation~~}  &\Longrightarrow {\rm ~~Universal~Extension~~} \\
\nonumber &\Longrightarrow  {\rm ~~Dynamically~Faithful~States~~}\\
&\Longrightarrow   {\rm ~~Dynamically~Faithful~Systems.}   
\end{align*}
In short, the possibility of conclusive teleportation guarantees the feasibility of process tomography, even in theories that violate  Local Tomography.

 \subsubsection{Purification}\label{subsec:purification}  
 
 A {\em purification} of a given state    is an extension of that state to a pure state of a composite system \cite{chiribella2010probabilistic,chiribella2014dilation,chiribella2015conservation,chiribella2016quantum,dariano2017quantum}.      { In the standard scenario where the state space of physical systems is a convex set, pure states are defined as the extreme points of the convex set.   A  diagrammatic definition of pure state, valid also  for theories where the scalars are not real numbers, was provided in Ref. \cite{chiribella2014distinguishability}.  The second definition is equivalent to the first in   every theory satisfying   Causality \cite{chiribella2010probabilistic,coecke2010causal,chiribella2011informational,chiribella2012quantum,coecke2013causal,chiribella2014dilation,chiribella2016quantum,dariano2017quantum,coecke2018picturing}, the requirement that  it should be  impossible to ``send signals from the future to the past''  (see below for a more precise mathematical statement) and Local Tomography, or  Local Tomography on Pure States (a weakening  of Local Tomography satisfied by quantum theory on real vector spaces\footnote{Specifically, Local Tomography on Pure States is the requirement that  Eq. \eqref{equalityLT}  implies Eq. \eqref{LT2}  when at least one of the two states $\rho$ and $\rho'$ is an extreme point of the state space.}).  

In the following, we will not commit to a specific definition of pure state.  Instead, we will simply assume that for every system $\rA$ there exists a set of states  $\Pur\St (\rA)$, called {\em pure},  satisfying the condition 
\begin{align}\label{pureproduct}
\alpha \in  \Pur\St (\rA)  \, ,  \,   \beta \in  \Pur\St (\rB)  \qquad  \Longrightarrow  \qquad  \alpha\otimes \beta \in  \Pur\St(\rA\otimes \rB)\, . 
\end{align}
We call the above condition Pure Product States.   Pure Product States is automatically  satisfied by the diagrammatic definition of pure state in  \cite{chiribella2014distinguishability}.    It  is also satisfied  by the usual definition of pure states as extreme points, when it coincides with the diagrammatic  definition.

 Let us now  review the notion of purification.   In its basic form, purification  is  formulated in  theories satisfying   Causality. 
An equivalent condition for Causality is the existence of a unique deterministic effect: 
\begin{prop}[\cite{chiribella2010probabilistic}]\label{prop:uniquedet}
A theory satisfies Causality if and only if for every system $\rA$ there exists a unique deterministic effect $u_\rA \in  \Det\Eff (\rA)$. 
\end{prop}

Theories satisfying Causality are often called {\em causal theories}.  In a causal theory,  a state $\Gamma \in  \St  ( \rA\otimes \rE )$ is an extension of  another state $\rho  \in  \St (\rA)$ if the condition
 	\begin{align}
	\begin{aligned} \Qcircuit @C=1em @R=.7em @!R {  
	  &   \multiprepareC{1}{\Gamma} & \qw \poloFantasmaCn{\rA} & \qw       & &=&   & \prepareC{\rho } & \qw \poloFantasmaCn{\rA} &\qw 
 \\ 
	& \pureghost{\Gamma} & \qw \poloFantasmaCn{\rE} & \measureD{u_{\rm E}} &  & &  &&  & }  \end{aligned}	\end{align} 
is satisfied. 

We are now ready to state the Purification principle in its basic form:  
\begin{prin}[Purification]
For a theory satisfying Causality and Pure Product States,  we require that
\begin{enumerate}
\item every state has a purification: for every system $\rA \in \Sys$ and every deterministic state $\rho  \in \Det\St (\rA)$ there exists a system $\rR$ and a deterministic pure state $\Psi  \in { \Pur\St (\rA\otimes \rR)}$ such that 
	\begin{align}
	\begin{aligned} \Qcircuit @C=1em @R=.7em @!R {  
	  &   \multiprepareC{1}{\Psi} & \qw \poloFantasmaCn{\rA} & \qw       & &=&   & \prepareC{\rho } & \qw \poloFantasmaCn{\rA} &\qw 
 \\ 
	& \pureghost{\Psi} & \qw \poloFantasmaCn{\rR} & \measureD{u_{\rm R}} &  & &  &&  &}   \end{aligned}	\end{align} 
\item  every two purifications with the same purifying system are  interconvertible via  a local symmetry transformation:  for every system $\rR$, and every pair of deterministic pure states $\Psi$ and $\Psi'$ in $\Pur\St (\rA \otimes \rR)$,  the condition  
	\begin{align}
	\begin{aligned} \Qcircuit @C=1em @R=.7em @!R {  
	  &   \multiprepareC{1}{\Psi'} & \qw \poloFantasmaCn{\rA} & \qw     
 \\ 
	& \pureghost{\Psi'} & \qw \poloFantasmaCn{\rR} & \measureD{u_{\rm R}}} 
	 \end{aligned}	\quad    =  	\begin{aligned} \Qcircuit @C=1em @R=.7em @!R {  
	  &   \multiprepareC{1}{\Psi} & \qw \poloFantasmaCn{\rA} & \qw     
\\ 
	& \pureghost{\Psi} & \qw \poloFantasmaCn{\rR} & \measureD{u_{\rm R}}} 
	 \end{aligned}	 \end{align} 
	 implies that there exists a reversible transformation $\map U: \rR \to \rR$ such that 
	 	\begin{align}
	\begin{aligned} \Qcircuit @C=1em @R=.7em @!R {  
	  &   \multiprepareC{1}{\Psi'} & \qw \poloFantasmaCn{\rA} & \qw     
 \\ 
	& \pureghost{\Psi'} & \qw \poloFantasmaCn{\rR} & \qw} 
	 \end{aligned}	 \quad =  	\begin{aligned} \Qcircuit @C=1em @R=.7em @!R {  
	  &   \multiprepareC{1}{\Psi} & \qw \poloFantasmaCn{\rA} & \qw   &\qw &\qw     
 \\ 
	& \pureghost{\Psi} & \qw \poloFantasmaCn{\rR} & \gate{\map U}   &  \qw  \poloFantasmaCn{\rR} &\qw  } 
	 \end{aligned}	 \end{align} 
\end{enumerate} 
\end{prin}

Remarkably, the symmetry property of purifications implies that every purification is a universal extension: 
\begin{prop}\label{prop:puriisuniversal}
Let  $\rho \in \Det\St (\rA)$ be an arbitrary state  of an  arbitrary system $\rA$,  and let $\Gamma \in \Det\St (\rA \otimes \rE)$ be an arbitrary  extension of $\rho$.
 If the theory satisfies Causality, Pure Product States, and Purification, then the state $\Gamma$ can be generated as  
 	 	\begin{align}
	\begin{aligned} \Qcircuit @C=1em @R=.7em @!R {  
	  &   \multiprepareC{1}{\Gamma} & \qw \poloFantasmaCn{\rA} & \qw     
 \\ 
	& \pureghost{\Gamma} & \qw \poloFantasmaCn{\rE} & \qw  &\qquad} 
	 \end{aligned}	  =  	\begin{aligned} \Qcircuit @C=1em @R=.7em @!R {   
	&\qquad   &   \multiprepareC{1}{\Psi} & \qw \poloFantasmaCn{\rA} & \qw   &\qw &\qw     
 \\ 
	  &  & \pureghost{\Psi} & \qw \poloFantasmaCn{\rR} & \gate{\map T}   &  \qw  \poloFantasmaCn{\rE} &\qw  } 
	 \end{aligned} ~ ,	 \end{align} 
	 where $\Psi\in\Pur\St (\rA\otimes \rR)$
 is an arbitrary purification of $\rho$ and $\map T \in \Det\Transf(  \rR \to \rE)$ is a deterministic transformation.  \end{prop}
 
 The proof is provided in Appendix \ref{app:puriisuniversal}.}  The obvious consequence of Proposition \ref{prop:puriisuniversal} is the following
   \begin{cor}[A1,A3,A4]\label{cor:purification}
  If a  theory satisfies Causality, Pure Product States, and Purification, then it satisfies Universal Extension, and, in particular,  Dynamically Faithful States.
    \end{cor}
  \Proof Proposition \ref{prop:puriisuniversal} shows that  every purification of a given state $\rho$  is a universal extension,  in the sense of   Definition  \ref{def:uniext}.    In particular, every complete state has a universal extension (complete states exist by Assumption \ref{ass:complete}).  Hence,  Universal Extension holds.  Then, Theorem \ref{theo:uniext} guarantees that Dynamically Faithful States holds. \qed   
    
Summarizing, we proved the chain of implications: 
\begin{align*}
\nonumber {\rm Causality,~Pure~Product~States,~Purification~~}  &\Longrightarrow {\rm ~~Universal~Extension~~} \\
\nonumber &\Longrightarrow  {\rm ~~Dynamically~Faithful~States~~}\\
&\Longrightarrow   {\rm ~~Dynamically~Faithful~Systems.}   
\end{align*}

Note that this is the same chain of implications as in the previous subsection, except that ``Conclusive Teleportation'' is now replaced by ``Causality + Pure Product States +  Purification.''  A natural question is whether there are relations between Purification and Conclusive Teleportation. When the scalars are real numbers in the interval $[0,1]$,  it is known that Purification implies Conclusive Teleportation \cite{chiribella2010probabilistic,chiribella2011informational,chiribella2016quantum,dariano2017quantum}.  More generally, it is also possible to show that Purification implies Conclusive Teleportation in every theory that satisfies two additional axioms of Pure Sharpness and Purity Preservation \cite{chiribella2017microcanonical,chiribella2015conservation,chiribella2015operational}.     



\section{Conclusions}  

In this paper we  analyzed  the task of process tomography in general physical theories, exploring the requirement  that  physical processes should be identifiable by their action on a finite set of auxiliary systems/a finite set of input states.  Most of the paper focussed on the  requirement that physical processes can be identified by their action   on a single state, called dynamically faithful.     The existence of dynamically faithful states  is a broader condition than the usual principle of Local Tomography, and is satisfied in a number of variants of quantum theory, including quantum theory on real Hilbert spaces and a Fermionic version of quantum theory \cite{d2014fermionic,lugli2020fermionic}.  It is natural to conjecture that dynamically faithful states exist also in a recent extension of standard quantum theory that includes complex, real, and quaternionic Hilbert spaces in a single theory  \cite{barnum2020composites}.  A formal proof of this statement, however, is currently missing and remains as an interesting direction for future research.

One of our main results is that the existence of dynamically faithful states can be guaranteed by a simple physical condition, namely that complete states have a universal extension from which all the other extensions can be generated with non-zero probability. 
 Example of physical theories with this property are the set of theories where conclusive teleportation is possible, and  the set of theories satisfying the Causality, Pure Product States, and Purification.   For  every theory satisfying  Causality, Pure Product States, and Purification, process tomography can be achieved using a single bipartite state: specifically,  the purification of any complete state of a given system $A$  can be used to characterize all the processes with input  system $A$ and arbitrary output system $B$. 

An important observation is that the existence of dynamically faithful states  is independent of  the Causality axiom, that is, the requirement that the outcome probabilities of present experiments are independent of the choice of future experiments. This observation suggests that the possibility of characterizing physical processes may have a more primitive role than considerations of causality, and that one could adopt Dynamically Faithful States as a foundational principle for general physical theories where causality is emergent.

  \bigskip
  
 {\bf Acknowledgements.}  {I wish to thank Mauro D'Ariano for introducing me to  quantum tomography  and   to the problem of deriving quantum theory from physical principles.  The discussions we had during my PhD and early postdoc years have been the seeds for much of the ideas developed in this paper.  I also wish to acknowledge discussions with Paolo Perinotti and Carlo Maria Scandolo on the relations between purification and the existence of dynamically faithful states.  
}  {This work is supported by the Hong Kong Research Grant Council through grant 17300918 and though the Senior Research Fellowship Scheme SRFS2021-7S02, by the Croucher Foundation, by the John Templeton Foundation through grant 61466, The Quantum Information Structure of Spacetime (qiss.fr).  Research at the Perimeter Institute is supported by the Government of Canada through the Department of Innovation, Science and Economic Development Canada and by the Province of Ontario through the Ministry of Research, Innovation and Science. The opinions expressed in this publication are those of the authors and do not necessarily reflect the views of the John Templeton Foundation.}




\bibliographystyle{apsrev4-1}
\bibliography{generic}

 \appendix  

 \section{A converse of Proposition \ref{prop:processtomo}}\label{app:protomo}

 Bipartite states from which other states can be probabilistically generated were introduced  in  the early work by D'Ariano \cite{dariano2006missing,dariano2006how,dariano2007operational,maurobook}, where they were called  {\em preparationally faithful}.     Originally \cite{dariano2006missing,dariano2006how}, the term referred to states of a bipartite system $\rA \otimes \rB$ with the property that every state of $\rA$ can be probabilistically generated  by a physical transformation on $\rB$.  In later works \cite{dariano2007operational,maurobook}, the term  was used for  states with the property that every joint state of $\rA\otimes \rB$ can be probabilistically generated by a physical transformation on $\rB$.    Here we adopt the second definition and extend  it to {\em (1)}  transformations with different inputs and outputs, and  {\em (2)}  theories where the scalars are not necessarily real numbers:   

\begin{defi}\label{def:prepfaith}
A state $\Phi  \in  \St (\rA  \otimes \rS)$  is  {\em preparationally faithful} from $\rS$ to  $\rB$ if    
  every state  $\rho \in  \St  (\rA\otimes \rB )$ can be probabilistically obtained  as 
 \begin{align}\label{prepfaith}
p\,  \begin{aligned} \Qcircuit @C=1em @R=.7em @!R {
	 & \pureghost{\rho} & \qw \poloFantasmaCn{\rA} &  \qw  \\
	 & \multiprepareC{-1}{\rho} & \qw \poloFantasmaCn{\rB}   &\qw  } 
\end{aligned}     
	 \quad    =  \begin{aligned} \Qcircuit @C=1em @R=.7em @!R {
	 & \pureghost{\Phi} & \qw \poloFantasmaCn{\rA} &  \qw  &\qw & \qw   \\
	 & \multiprepareC{-1}{\Phi} & \qw \poloFantasmaCn{\rS}   &\gate{ \map S}  &    \qw \poloFantasmaCn{\rB}  & \qw  ~~~~\,,  } 
\end{aligned}  
	 \end{align}
where  $p\in\Transf(\rI\to \rI)$ is a cancellative scalar and $\map S$ is a transformation of type $\rS\to \rB$.  
\end{defi}  

Using the above definition, we introduce a notion of {\em doubly preparationally faithful theories}:  
\begin{defi}
A theory is {\em   doubly preparationally faithful} if for every pair of systems $\rA $ and $\rB$ there exists a system $\rS$ and a state $\Phi \in \St (\rA\otimes \rS)$ that is  preparationally faithful from $\rA$ to $\rA$ and from $\rS$ to $\rB$. 
\end{defi}
Concrete examples of doubly  preparationally faithful theories are finite-dimensional classical probability theory and quantum theory, both on complex and on real vector spaces. In quantum theory,   the maximally entangled state $|\Phi\>  =  \sum_{k=1}^{d_\rA} \,  |k\>\otimes |k\>/\sqrt d$ of system $\rA\otimes \rS$, with $\rS  \simeq  \rA$,   is preparationally faithful from $\rS$ to $\rB$ with arbitrary $\rB$.  Indeed, every pure state $|\Psi\>  =  \sum_{k=1}^{d_\rA} \sum_{l=1}^{d_\rB}  \,  c_{kl} \,  |k\>\otimes |l\>$ of the composite system $\rA \otimes \rB$ can be probabilistically obtained from  the maximally entangled state as $|\Psi\>  \propto   (     I_{\rA}\otimes   K ) |\Phi\>$, where $K  =   \sum_{k,l}  \, c_{kl}   |l\>\<k| /  \sqrt{\|  c^\dag  c\|}$ is a suitable Kraus operator, $c$ is the matrix with matrix elements $ c_{kl} $, and  $\|  c^\dag  c\|$ is the maximum eigenvalue of $c^\dag c$.   Since every pure state can be generated probabilistically from $|\Phi\>$, the same holds for mixed states (in finite dimensions). This shows that  $|\Phi\>$ is preparationally faithful from $\rS$ to $\rB$, with arbitrary $\rB$.   A similar argument shows that $|\Phi\>$ is preparationally faithful from $\rA$ to $\rA$.

More generally, every process  theory where the processes form a compact closed category \cite{abramsky2004,coecke2006kindergarten,abramsky2008,coecke2010quantum} is doubly preparationally faithful.  Physically, these are the theories where every system $\rA$ comes with a dual system $\rS$, such that  there exists a state of $\rA \otimes \rS$ that allows for conclusive teleportation of states of $\rA$, as well as conclusive teleportation of states of $\rS$.

With the above definitions, we have the following proposition:
\begin{prop}[A3]\label{prop:protomo}
For a  doubly preparationally faithful theory, the following are equivalent:  
\begin{enumerate} 
\item Local Tomography holds 
\item standard  process tomography completely identifies physical processes, that is,  Eq. (\ref{stan}) implies Eq. (\ref{sameprocess}). 
\end{enumerate}
\end{prop}

\Proof  Since the implication $1  \Longrightarrow 2$ was already proved by Proposition \ref{prop:processtomo}, we only need to prove  the implication $2\Longrightarrow  1$.   

Let $\rA$ and $\rB$ be two generic systems, and let $\rho$ and $\rho'$ be two generic states of the composite system $\rA\otimes \rB$.   Our goal is to show that then the condition  
\begin{align}\label{equalityLT1}
\begin{aligned} 
\Qcircuit @C=1em @R=.7em @!R {
	 & \multiprepareC{1}{\rho} & \qw \poloFantasmaCn{\rA}   & \measureD{a}  \\ 
	 & \pureghost{\rho} & \qw \poloFantasmaCn{\rB} &  \measureD{b} 
	 }
\end{aligned}    
\quad = 
\begin{aligned} 
\Qcircuit @C=1em @R=.7em @!R {
	 & \multiprepareC{1}{\rho'} & \qw \poloFantasmaCn{\rA}   & \measureD{a}   &\qquad  &\qquad&&\forall a\in\Eff (\rA)  &\qquad&\qquad & \\ 
	 & \pureghost{\rho'} & \qw \poloFantasmaCn{\rB} &  \measureD{b}   &\qquad &\qquad&& \forall b\in\Eff (\rB)  &\qquad&\qquad &
	 }
\end{aligned}    
\end{align}
implies $  \rho  =  \rho'$.   

Let us assume that Eq. (\ref{equalityLT1}) holds.  Since the theory is doubly preparationally faithful, there exists a system $\rS$ and a state $\Phi \in \St \left(\rA\otimes \rS\right)$ that is preparationally faithful  from $\rA$ to $\rA$ and from $\rS$ to $\rB$.   Since $\Phi$ is preparationally faithful from $\rS$ to $\rB$,  there exist   two cancellative scalars $p$ and $p'$, and two transformations  $\map S$ and $\map S'$, of type $\rS\to \rB$, such that  
\begin{align}\label{sclero1}
p\,  \begin{aligned} \Qcircuit @C=1em @R=.7em @!R {
	 & \pureghost{\rho} & \qw \poloFantasmaCn{\rA} &  \qw  \\
	 & \multiprepareC{-1}{\rho} & \qw \poloFantasmaCn{\rB}   &\qw  } 
\end{aligned}     
	 \quad    =  \begin{aligned} \Qcircuit @C=1em @R=.7em @!R {
	 & \pureghost{\Phi} & \qw \poloFantasmaCn{\rA} &  \qw  &\qw & \qw   \\
	 & \multiprepareC{-1}{\Phi} & \qw \poloFantasmaCn{\rS}   &\gate{  \map S}  &    \qw \poloFantasmaCn{\rB}  & \qw  ~~~~~\,,  } 
\end{aligned}  
\end{align}  
and 
\begin{align}\label{sclero2}
p'\,  \begin{aligned} \Qcircuit @C=1em @R=.7em @!R {
	 & \pureghost{\rho'} & \qw \poloFantasmaCn{\rA} &  \qw  \\
	 & \multiprepareC{-1}{\rho'} & \qw \poloFantasmaCn{\rB}   &\qw  } 
\end{aligned}     
	 \quad    =  \begin{aligned} \Qcircuit @C=1em @R=.7em @!R {
	 & \pureghost{\Phi} & \qw \poloFantasmaCn{\rA} &  \qw  &\qw & \qw   \\
	 & \multiprepareC{-1}{\Phi} & \qw \poloFantasmaCn{\rS}   &\gate{ \map S'}  &    \qw \poloFantasmaCn{\rB}  & \qw  ~~~~~\,.  } 
\end{aligned}  
\end{align}   

Since the state $\Phi$ is preparationally faithful  from  $\rA$ to $\rA$,  for every state $\Sigma \in \St (\rA\otimes \rS)$ there exists a cancellative scalar $q$ and  transformation $\map T$ of type $\rA \to \rA$ such that 
\begin{align}
q\,  \begin{aligned} \Qcircuit @C=1em @R=.7em @!R {
	 & \pureghost{\Sigma} & \qw \poloFantasmaCn{\rA} &  \qw  \\
	 & \multiprepareC{-1}{\Sigma} & \qw \poloFantasmaCn{\rS}   &\qw  } 
\end{aligned}     
	 \quad    =  \begin{aligned} \Qcircuit @C=1em @R=.7em @!R {
	 & \pureghost{\Phi} & \qw \poloFantasmaCn{\rA}   &\gate{  \map T}  &    \qw \poloFantasmaCn{\rA}  & \qw 
	   \\
	 & \multiprepareC{-1}{\Phi}   & \qw \poloFantasmaCn{\rS} &  \qw  &\qw & \qw ~~~~~ .  } 
\end{aligned}  
\end{align}  
 Inserting this relation into Eqs. (\ref{sclero1}) and (\ref{sclero2}), we obtain 
\begin{align}
p\,  \begin{aligned} \Qcircuit @C=1em @R=.7em @!R {
	 & \pureghost{\rho} & \qw \poloFantasmaCn{\rA} &  \gate{\map T}  &  \qw \poloFantasmaCn{\rA}  & \qw \\
	 & \multiprepareC{-1}{\rho} & \qw \poloFantasmaCn{\rB}   &\qw   &\qw &\qw  } 
\end{aligned}     
	 \quad    = q\,   \begin{aligned} \Qcircuit @C=1em @R=.7em @!R {
	 & \pureghost{\Sigma} & \qw \poloFantasmaCn{\rA} &  \qw  &\qw & \qw   \\
	 & \multiprepareC{-1}{\Sigma} & \qw \poloFantasmaCn{\rS}   &\gate{  \map S}  &    \qw \poloFantasmaCn{\rB}  & \qw  ~~~~~\,,  } 
\end{aligned}  
\end{align}  
and 
\begin{align}
p'\,  \begin{aligned} \Qcircuit @C=1em @R=.7em @!R {
	 & \pureghost{\rho'} & \qw \poloFantasmaCn{\rA} &  \gate{\map T}  &  \qw \poloFantasmaCn{\rA}  & \qw \\
	 & \multiprepareC{-1}{\rho'} & \qw \poloFantasmaCn{\rB}   &\qw   &\qw &\qw  } 
\end{aligned}     
	 \quad    = q\,   \begin{aligned} \Qcircuit @C=1em @R=.7em @!R {
	 & \pureghost{\Sigma} & \qw \poloFantasmaCn{\rA} &  \qw  &\qw & \qw   \\
	 & \multiprepareC{-1}{\Sigma} & \qw \poloFantasmaCn{\rS}   &\gate{  \map S'}  &    \qw \poloFantasmaCn{\rB}  & \qw  ~~~~~\,,  } 
\end{aligned}  
\end{align}  

Now, note that  $p'  \,\map S$ and $p\,  \map S'$ are physical transformations, due to the Randomizations Assumption \ref{ass:random}.  
For these two transformations, we have the relation 
 \begin{align}
\nonumber q\,     \begin{aligned} \Qcircuit @C=1em @R=.7em @!R {
	 & \pureghost{\Sigma} & \qw \poloFantasmaCn{\rA} &  \qw  &\qw & \measureD{a'}   \\
	 & \multiprepareC{-1}{\Sigma} & \qw \poloFantasmaCn{\rS}   &\gate{p'\,   \map S}  &    \qw \poloFantasmaCn{\rB}  & \measureD{b} ~~~~~\,,  } 
\end{aligned}    
  &=    \quad  p\,  p'\,        \begin{aligned} \Qcircuit @C=1em @R=.7em @!R {
	 & \pureghost{\rho} & \qw \poloFantasmaCn{\rA} & \gate{\map T}  & \qw \poloFantasmaCn{\rA}  &  \measureD{a'} \\
	 & \multiprepareC{-1}{\rho} & \qw \poloFantasmaCn{\rB}    &\qw &\qw &\measureD{b}  } 
\end{aligned}   \\
\nonumber    & \\
\nonumber    &=    \quad  p\,  p'\,        \begin{aligned} \Qcircuit @C=1em @R=.7em @!R {
	 & \pureghost{\rho'} & \qw \poloFantasmaCn{\rA} & \gate{\map T}  & \qw \poloFantasmaCn{\rA}  &  \measureD{a'} \\
	 & \multiprepareC{-1}{\rho'} & \qw \poloFantasmaCn{\rB}    &\qw &\qw &\measureD{b}  } 
\end{aligned}   \\ 
\nonumber & \\
   &  
   = \quad  q\,     \begin{aligned} \Qcircuit @C=1em @R=.7em @!R {
	 & \pureghost{\Sigma} & \qw \poloFantasmaCn{\rA} &  \qw  &\qw & \measureD{a'}    &\qquad  &\qquad&&\forall a'\in\Eff (\rA)  &\qquad&\qquad &  \\
	 & \multiprepareC{-1}{\Sigma} & \qw \poloFantasmaCn{\rS}   &\gate{p\,   \map S'}  &    \qw \poloFantasmaCn{\rB}  & \measureD{b}  &\qquad  &\qquad&&\forall b\in\Eff (\rB)  &\qquad&\qquad & ~~\, ,   } 
\end{aligned} \label{neardeathexperience} 
\end{align} 
where the second equality follows from Eq. (\ref{equalityLT1}), applied to 
\begin{align}
   \begin{aligned} \Qcircuit @C=1em @R=.7em @!R {   & \qw \poloFantasmaCn{\rA} &  \measureD{a} } 
   \end{aligned}\quad  :  = \quad     \begin{aligned} \Qcircuit @C=1em @R=.7em @!R {   & \qw \poloFantasmaCn{\rA} &\gate{\map T}  & \qw \poloFantasmaCn{\rA}   & \measureD{a'} } ~~.
   \end{aligned}
   \end{align}

Recall that $\Sigma$ is a generic state of system $\rA\otimes \rS$. In particular, we can choose $\Sigma$ to be of the product form  $\Sigma  = \alpha\otimes \sigma$. 
  With this choice, Eq. \eqref{neardeathexperience} becomes  
   \begin{align}
q\,     \begin{aligned} \Qcircuit @C=1em @R=.7em @!R {
	 & \prepareC{\alpha} & \qw \poloFantasmaCn{\rA} &  \qw  &\qw & \measureD{a'}   \\
	 & \prepareC{\sigma} & \qw \poloFantasmaCn{\rS}   &\gate{p'\,   \map S}  &    \qw \poloFantasmaCn{\rB}  & \measureD{b}   } 
\end{aligned}    
  &=  \quad  q\,     \begin{aligned} \Qcircuit @C=1em @R=.7em @!R {
	 & \prepareC{\alpha} & \qw \poloFantasmaCn{\rA} &  \qw  &\qw & \measureD{a'}    &\qquad  &\qquad&&\forall a'\in\Eff (\rA)  &\qquad&\qquad &  \\
	 & \prepareC{\sigma} & \qw \poloFantasmaCn{\rS}   &\gate{p\,   \map S'}  &    \qw \poloFantasmaCn{\rB}  & \measureD{b}  &\qquad  &\qquad&&\forall b\in\Eff (\rB)  &\qquad&\quad & \, .  }   \label{ultimominuto}
\end{aligned} 
\end{align} 
Choosing $\alpha$ to be a deterministic state and $a'$ to be a deterministic effect, we have 
\begin{align}
 \begin{aligned} \Qcircuit @C=1em @R=.7em @!R {
	 & \prepareC{\alpha} & \qw \poloFantasmaCn{\rA}  & \measureD{a'}     &  =    & 1 }
\end{aligned}
\end{align}
and Eq. \eqref{ultimominuto} becomes  
  \begin{align}
q\,     \begin{aligned} \Qcircuit @C=1em @R=.7em @!R {
	 & \prepareC{\sigma} & \qw \poloFantasmaCn{\rS}   &\gate{p'\,   \map S}  &    \qw \poloFantasmaCn{\rB}  & \measureD{b}  } 
\end{aligned}    
  &=  \quad  q\,     \begin{aligned} \Qcircuit @C=1em @R=.7em @!R {
	 & \prepareC{\sigma} & \qw \poloFantasmaCn{\rS}   &\gate{p\,   \map S'}  &    \qw \poloFantasmaCn{\rB}  & \measureD{b}  &\qquad  &\qquad&&\forall b\in\Eff (\rB) & \qquad  &\quad & \, .    } 
\end{aligned} 
\end{align} 
Since $q$ is cancellative for every $\alpha$ and $\sigma$,  we then obtain
 \begin{align}
    \begin{aligned} \Qcircuit @C=1em @R=.7em @!R {
	 & \prepareC{\sigma} & \qw \poloFantasmaCn{\rS}   &\gate{p'\,   \map S}  &    \qw \poloFantasmaCn{\rB}  & \measureD{b}  } 
\end{aligned}    
  &=       \begin{aligned} \Qcircuit @C=1em @R=.7em @!R {
	 & \prepareC{\sigma} & \qw \poloFantasmaCn{\rS}   &\gate{p\,   \map S'}  &    \qw \poloFantasmaCn{\rB}  & \measureD{b}  &\qquad  &\qquad&&\forall b\in\Eff (\rB)   &\qquad &\quad  &\, .   } 
\end{aligned} 
\end{align}  
To conclude, recall that $\sigma$ is an arbitrary state, and therefore the above equation implies that the processes $p\, \map S'$ and $p'\, \map S$ give rise to the same outcome probability in every setup of standard process tomography.  Since by hypothesis standard process tomography  successfully identifies physical processes, we conclude that the equality $p\,  \map S'  =  p'\,\map S$ must hold.  
   Using this fact and  Eqs. (\ref{sclero1}) and (\ref{sclero2}),we obtain 
   \begin{align}
\nonumber     p\,p'\,    \begin{aligned} \Qcircuit @C=1em @R=.7em @!R {
	 & \pureghost{\rho} & \qw \poloFantasmaCn{\rA} &  \qw  \\
	 & \multiprepareC{-1}{\rho} & \qw \poloFantasmaCn{\rB}   &\qw  } 
\end{aligned}     
	 \quad    &= \quad  
	 \begin{aligned} \Qcircuit @C=1em @R=.7em @!R {
	 & \pureghost{\Phi} & \qw \poloFantasmaCn{\rA} &  \qw  &\qw & \qw   \\
	 & \multiprepareC{-1}{\Phi} & \qw \poloFantasmaCn{\rS}   &\gate{ p' \,   \map S}  &    \qw \poloFantasmaCn{\rB}  & \qw   } 
\end{aligned}    \\
\nonumber	 &\\
\nonumber	 &  = \quad  \begin{aligned} \Qcircuit @C=1em @R=.7em @!R {
	 & \pureghost{\Phi} & \qw \poloFantasmaCn{\rA} &  \qw  &\qw & \qw   \\
	 & \multiprepareC{-1}{\Phi} & \qw \poloFantasmaCn{\rS}   &\gate{ p \,   \map S'}  &    \qw \poloFantasmaCn{\rB}  & \qw   }
\end{aligned}  \\	 
\nonumber	 &\\
	 &  = \quad    p\,p'\,    \begin{aligned} \Qcircuit @C=1em @R=.7em @!R {
	 & \pureghost{\rho'} & \qw \poloFantasmaCn{\rA} &  \qw  \\
	 & \multiprepareC{-1}{\rho'} & \qw \poloFantasmaCn{\rB}   &\qw  } 
\end{aligned}    \, .
\end{align}
 Since $p$ and $p'$ are both cancellative, we conclude that the equality $\rho  = \rho'$must hold.  \qed

 \section{Properties of the relations $\equiv_\rho$ and $=_\rho$}\label{app:LT}

We start by showing that the relation $\equiv_\rho$ is  stronger than the relation $=_\rho$.
\medskip  

{\bf Proof of Lemma \ref{lem:intermediate}.}    Let $\bs \rho  = (\rho_x)_{x\in\set X}$ be a source with average state $\rho$.   The Display Assumption \ref{ass:display} guarantees that  the states in $\bs \rho$ can be prepared  as 
	\begin{align}\label{rhoxdisplay}
	\begin{aligned} \Qcircuit @C=1em @R=.7em @!R {  
	& \prepareC{\rho_{x} } & \qw \poloFantasmaCn{\rA} &\qw     & &=&   & \multiprepareC{1}{\Gamma} & \qw \poloFantasmaCn{\rA} & \qw   &   \qquad   &\forall x\in\mathsf{X} 
 \\ 
	&&&&  &&  & \pureghost{\Gamma} & \qw \poloFantasmaCn{\rE} & \measureD{e_x}} &  & \end{aligned}	\end{align} 
for some system $\mathrm{\rE}$, some  deterministic state $\Gamma $, and some measurement  $\su e  =  (e_x)_{x\in\set X}$. 
 Now, the state $\Gamma$ is an extension of $\rho$: indeed,  one has 
  	\begin{align}
\nonumber 	\begin{aligned} \Qcircuit @C=1em @R=.7em @!R {  
	& \prepareC{\rho} & \qw \poloFantasmaCn{\rA} &\qw } 
	\end{aligned} \quad   &= \quad   \sum_{x\in\set X}   
	\begin{aligned} \Qcircuit @C=1em @R=.7em @!R {  
	&   \prepareC{\rho_x} & \qw \poloFantasmaCn{\rA} & \qw}  \end{aligned}\\ 
	\nonumber & \\
 \nonumber  & = \quad   \sum_{x\in\set X}   
	\begin{aligned} \Qcircuit @C=1em @R=.7em @!R {  
	&   \multiprepareC{1}{\Gamma} & \qw \poloFantasmaCn{\rA} & \qw \\ 
  & \pureghost{\Gamma} & \qw \poloFantasmaCn{\rE} & \measureD{e_x}}  \end{aligned}\\
  \nonumber &  \\
  &   =  	\begin{aligned} \Qcircuit @C=1em @R=.7em @!R {  
	&   \multiprepareC{1}{\Gamma} & \qw \poloFantasmaCn{\rA} & \qw \\ 
  & \pureghost{\Gamma} & \qw \poloFantasmaCn{\rE} & \measureD{e}}   \end{aligned} \quad  ,  	\end{align}
  where $e$ is the deterministic effect $e:=  \sum_{x\in\set X}  e_x$ (the last equality follows from Eq. (\ref{coarse3})).  
   
 Now, suppose that the condition  
 \begin{align}
\begin{aligned} \Qcircuit @C=1em @R=.7em @!R { & \multiprepareC{1}{\Gamma} & \qw \poloFantasmaCn{\rA} &  \gate{\map P}  &  \qw \poloFantasmaCn{\rB} & \qw 
 \\ 
	 & \pureghost{\Gamma} & \qw \poloFantasmaCn{\rE} &  \qw &\qw   &\qw} \end{aligned} \quad   =  
	 \begin{aligned} \Qcircuit @C=1em @R=.7em @!R { & \multiprepareC{1}{\Gamma} & \qw \poloFantasmaCn{\rA} &  \gate{\map P'}  &  \qw \poloFantasmaCn{\rB} & \qw  
 \\ 
	 & \pureghost{\Gamma} & \qw \poloFantasmaCn{\rE} &  \qw &\qw   &  \qw} \end{aligned}	 
	 \end{align}
	 holds.   Applying the effect $e_x$ on both sides, one then obtains 
 \begin{align}
\begin{aligned} \Qcircuit @C=1em @R=.7em @!R { & \multiprepareC{1}{\Gamma} & \qw \poloFantasmaCn{\rA} &  \gate{\map P}  &  \qw \poloFantasmaCn{\rB} & \qw 
 \\ 
	 & \pureghost{\Gamma} & \qw \poloFantasmaCn{\rE} &  \qw   &\measureD{e_x}} \end{aligned} \quad   =  
	 \begin{aligned} \Qcircuit @C=1em @R=.7em @!R { & \multiprepareC{1}{\Gamma} & \qw \poloFantasmaCn{\rA} &  \gate{\map P'}  &  \qw \poloFantasmaCn{\rB} & \qw  
 \\ 
	 & \pureghost{\Gamma} & \qw \poloFantasmaCn{\rE} &  \qw   &  \measureD{e_x}} \end{aligned}	
	  \end{align}
   and, using Eq. (\ref{rhoxdisplay}), 
  \begin{align}
\begin{aligned}
 \Qcircuit @C=1em @R=.7em @!R { & \prepareC{\rho_x} & \qw \poloFantasmaCn{\rA} &  \gate{\map P}  &  \qw \poloFantasmaCn{\rB} & \qw }
\end{aligned}
\quad =  
\begin{aligned}
 \Qcircuit @C=1em @R=.7em @!R { & \prepareC{\rho_x} & \qw \poloFantasmaCn{\rA} &  \gate{\map P'}  &  \qw \poloFantasmaCn{\rB} & \qw }  \qquad \forall x\in \set X\, .
\end{aligned}
 \end{align}
\qed 
\medskip 


We now show that the two relations  $\equiv_\rho$ and $=_\rho$ are equivalent in theories satisfying Causality and Local Tomography. We recall that Causality is equivalent to the existence of a unique deterministic effect  $u_\rA$ for every system $\rA$   \cite{chiribella2010probabilistic,coecke2010causal,coecke2013causal,chiribella2014dilation,chiribella2016quantum}.

We can then prove the following proposition:
 \begin{prop}[A1]
 Suppose that a theory satisfies  Causality and Local Tomography. Then, for every system $\rA$, every deterministic state $\rho\in\Det\St(\rA)$, and every pair of processes $\map P$ and $\map P'$, of type $\rA\to \rB$, one has the condition
 \begin{align}
 \map P  \equiv_{\rho}  \map P'  \qquad \Longleftrightarrow \qquad \map P  =_\rho  \map P' \, .  
 \end{align}
 \end{prop}

 \Proof We know from Corollary \ref{cor:deterministic} that the condition $\map P  \equiv_\rho \map P'$ implies the condition $\map P    =_\rho \map P'$.  To prove the equivalence, we only need to show the converse implication.  
Suppose that the condition  $\map P    =_\rho \map P'$ holds. Then, let $\Gamma$ be a generic extension of $\rho$ on system $\rE$, namely
\begin{align}\label{rhoext}
\begin{aligned} 
\Qcircuit @C=1em @R=.7em @!R {
	 & \prepareC{\rho} & \qw \poloFantasmaCn{\rA}   &\qw}
\end{aligned}    
\quad =  \begin{aligned} 
\Qcircuit @C=1em @R=.7em @!R {
	 & \multiprepareC{1}{\Gamma} & \qw \poloFantasmaCn{\rA}   & \qw  \\ 
	 & \pureghost{\Gamma} & \qw \poloFantasmaCn{\rE} &\measureD{u_\rE} 
	 }
\end{aligned}    ~.
\end{align}
Here we used the Causality Axiom, which guarantees that there exists only one deterministic effect for system $\rE$. 

At this point,  we need to prove the relation
  \begin{align}\label{toprove}
\begin{aligned} 
\Qcircuit @C=1em @R=.7em @!R {
	 & \multiprepareC{1}{\Gamma} & \qw \poloFantasmaCn{\rA}   & \gate{\map P}  &  \qw \poloFantasmaCn{\rB}  & \qw  \\ 
	 & \pureghost{\Gamma} & \qw \poloFantasmaCn{\rE} &  \qw  &\qw &\qw 
	 }
\end{aligned}    
\quad = \quad 
\begin{aligned} 
\Qcircuit @C=1em @R=.7em @!R {
	 & \multiprepareC{1}{\Gamma} & \qw \poloFantasmaCn{\rA}   & \gate{\map P}  &  \qw \poloFantasmaCn{\rB}  & \qw  \\ 
	 & \pureghost{\Gamma} & \qw \poloFantasmaCn{\rE} &  \qw  &\qw &\qw 
	 } 
\end{aligned}    ~. 
\end{align}
To this purpose, pick an arbitrary measurement $\su e  =  (e_x)_{x\in\set X}$ and define the preparation test $\bs \rho  =  (\rho_x)_{x\in\set X}$ 
as 
  \begin{align}\label{defrhox}
\begin{aligned} 
\Qcircuit @C=1em @R=.7em @!R {
	 & \prepareC{\rho_x} & \qw \poloFantasmaCn{\rA}   &\qw}
\end{aligned}    
\quad :=  \begin{aligned} 
\Qcircuit @C=1em @R=.7em @!R {
	 & \multiprepareC{1}{\Gamma} & \qw \poloFantasmaCn{\rA}   & \qw  \\ 
	 & \pureghost{\Gamma} & \qw \poloFantasmaCn{\rE} &\measureD{e_x} 
	 }
\end{aligned}    ~. 
\end{align}
Note that the preparation test $\bs \rho$ has average state $\rho$:  indeed,  one has 
   \begin{align}
\nonumber \sum_{x\in\set X}
\begin{aligned} 
\Qcircuit @C=1em @R=.7em @!R {
	 & \prepareC{\rho_x} & \qw \poloFantasmaCn{\rA}   &\qw}
\end{aligned}    
\quad &=  \begin{aligned} 
\Qcircuit @C=1em @R=.7em @!R {
	 & \multiprepareC{1}{\Gamma} & \qw \poloFantasmaCn{\rA}   & \qw  \\ 
	 & \pureghost{\Gamma} & \qw \poloFantasmaCn{\rE} &\measureD{u_\rE} 
	 }
\end{aligned}\\
\nonumber &  \\
&  =     
\begin{aligned} 
\Qcircuit @C=1em @R=.7em @!R {
	 & \prepareC{\rho} & \qw \poloFantasmaCn{\rA}   &\qw}
\end{aligned}     ~,
\end{align}
having used  Eq. (\ref{rhoext})    and the fact that $u_A$ is the only deterministic effect on system $\rA$.     

Since the preparation test $\bs \rho$ averages to $\rho$, the condition $\map P  = _\rho \map P'$ implies  
 \begin{align}
\begin{aligned} 
\Qcircuit @C=1em @R=.7em @!R {
	 & \prepareC{\rho_x} & \qw \poloFantasmaCn{\rA}   &\gate{\map P}  & \qw \poloFantasmaCn{\rB}  & \qw}
\end{aligned}    
\quad = 
\begin{aligned}
\Qcircuit @C=1em @R=.7em @!R {
	 & \prepareC{\rho_x} & \qw \poloFantasmaCn{\rA}   &\gate{\map P'}  & \qw \poloFantasmaCn{\rB}  & \qw  &  \qquad &  \qquad & \forall x\in\set X}
\end{aligned}  
\end{align}
and, using the definition of $\rho_x$ in Eq. (\ref{defrhox}),   
\begin{align}
\begin{aligned}
\Qcircuit @C=1em @R=.7em @!R {
	 & \multiprepareC{1}{\Gamma} & \qw \poloFantasmaCn{\rA}     &\gate{\map P}  & \qw \poloFantasmaCn{\rB} & \qw  \\ 
	 & \pureghost{\Gamma} & \qw \poloFantasmaCn{\rE} & \qw   &\measureD{e_x} & 
	 }
\end{aligned}    \quad
=
\begin{aligned}
\Qcircuit @C=1em @R=.7em @!R {
	 & \multiprepareC{1}{\Gamma} & \qw \poloFantasmaCn{\rA}     &\gate{\map P}  & \qw \poloFantasmaCn{\rB} & \qw  &&&\\ 
	 & \pureghost{\Gamma} & \qw \poloFantasmaCn{\rE} & \qw   &\measureD{e_x}    & \qquad &  \qquad & \forall x\in\set X &\qquad. 
	 }
\end{aligned}    
\end{align}
    To conclude, note that for a generic measurement $\su b  =  (b_y)_{y\in\set Y}$ one has 
\begin{align}
\begin{aligned}
\Qcircuit @C=1em @R=.7em @!R {
	 & \multiprepareC{1}{\Gamma} & \qw \poloFantasmaCn{\rA}     &\gate{\map P}  & \qw \poloFantasmaCn{\rB} & \measureD{b_y}  \\ 
	 & \pureghost{\Gamma} & \qw \poloFantasmaCn{\rE} & \qw   &\qw &\measureD{e_x} 
	 }
\end{aligned}    \quad
=
\begin{aligned}
\Qcircuit @C=1em @R=.7em @!R {
	 & \multiprepareC{1}{\Gamma} & \qw \poloFantasmaCn{\rA}     &\gate{\map P}  & \qw \poloFantasmaCn{\rB} & \measureD{b_y}  \\ 
	 & \pureghost{\Gamma} & \qw \poloFantasmaCn{\rE} & \qw    &\qw &\measureD{e_x}   
	 }
\end{aligned}    
\end{align}
for every $x\in\set X$ and for every $y\in\set Y$. Since the two measurements $\su b$ and $\su e$ are generic, Local Tomography  implies Eq. (\ref{toprove}). Since $\Gamma$ is a generic extension of $\rho$, we conclude that $\map P  \equiv_\rho  \map P'$.   \qed

\section{Proof of Lemma \ref{lem:equalonextensionsofdeterministic}}\label{app:extensiondet}  

 The direction $\Rightarrow$ is immediate.  To prove the direction $\Leftarrow$, let us assume the condition $\map P\equiv_\rho \map P'$ for every deterministic state $\rho$. Recall that two processes $\map P$ and $\map P'$ are equal if and only if
\begin{align}\label{onxi}
\begin{aligned} \Qcircuit @C=1em @R=.7em @!R { & \multiprepareC{1}{\Gamma} & \qw \poloFantasmaCn{\rA} &  \gate{\map P}  &  \qw \poloFantasmaCn{\rB} & \qw 
 \\ 
	 & \pureghost{\Gamma} & \qw \poloFantasmaCn{\rE} &  \qw &\qw   &\qw} \end{aligned} \quad   =  
	 \begin{aligned} \Qcircuit @C=1em @R=.7em @!R { & \multiprepareC{1}{\Gamma} & \qw \poloFantasmaCn{\rA} &  \gate{\map P'}  &  \qw \poloFantasmaCn{\rB} & \qw  
 \\ 
	 & \pureghost{\Gamma} & \qw \poloFantasmaCn{\rE} &  \qw &\qw   &  \qw} \end{aligned}	
	 \end{align}
	 for every system $\rE$ and for every (possibly non-deterministic) state $\Gamma\in\St (\rA\otimes \rE)$. 
	 
	  Let us pick a generic $\rE$ and a generic $\Gamma$.  Since every state can be prepared in some test (by the Displays Assumption \ref{ass:display}),  there must exist a test $\bs \Delta   =   (\Delta_x)_{x\in\set X}$  and an outcome $x_0\in \set X$   such that 
 \begin{align}\label{xi}
  \Delta_{x_0}   =    \Gamma  \,  .
 \end{align}
 Now, let us define the deterministic states 
 \begin{align}
 \Delta:  = \sum_{x\in\set X}  \, \Delta_x  
 \end{align}
 and 
  	\begin{align}
	\begin{aligned} \Qcircuit @C=1em @R=.7em @!R {  
  & \prepareC{\rho } & \qw \poloFantasmaCn{\rA} &\qw   	   & &:=&    &  \multiprepareC{1}{\Delta} & \qw \poloFantasmaCn{\rA} & \qw 
 \\ 
	&  & &&& && \pureghost{\Delta} & \qw \poloFantasmaCn{\rE} & \measureD{e}}   
	\end{aligned}	\end{align} 
where $e $ is a fixed, but otherwise arbitrary,  deterministic effect on system $\rE$.   By construction, $\Delta$ is an extension of $\rho$. Hence, every extension of $\Delta$ is also an extension of $\rho$. As a consequence, one has  the implication 
 \begin{align}
 \map P  \equiv_\rho  \map P'  \qquad \Longrightarrow \qquad \map P\otimes \map I_E  \equiv_{\Delta}   \,  \map P'  \otimes \map I_E   \, .
 \end{align} 
 Moreover, one has the implications  
 \begin{align}
 \map P \otimes \map I_E    \equiv_{\Delta}   \,  \map P'    \otimes \map I_E     \qquad  & \Longrightarrow \qquad   \map P \otimes \map I_E    =_{\Delta}   \,  \map P'   \otimes \map I_E  \end{align}
  (by Corollary \ref{cor:deterministic}) and 
  \begin{align}
\nonumber  \map P  \otimes \map I_E   =_{ \Delta}   \,  \map P'  \otimes \map I_E  \qquad   \Longrightarrow \qquad      &(\map P\otimes \map I_\rE  ) \Delta_x  =     (\map P'\otimes \map I_\rE  ) \Delta_x  \, ,    \end{align}
for every outcome $x$  (by definition of equality upon input of $\Delta$).  Choosing  $x=  x_0$ and using Eq. (\ref{xi}), we finally obtain  Eq. (\ref{onxi}).   Since  the system $\rE$ and the state $\Gamma$  are generic, the equality  $\map P= \map P'$ follows.     \qed

\section{Proof of Lemma \ref{lem:containtomo}}\label{app:containtomo}

Suppose that $\rho$ is a mixture  of $\sigma$ and $\tau$ with probabilities $(p_0,p_1)$, namely  
\begin{align}
\begin{aligned} 
\Qcircuit @C=1em @R=.7em @!R {  
& \prepareC{\rho } & \qw \poloFantasmaCn{\rA} &\qw  &\qquad   }  
\end{aligned}   
    =   \qquad p_0     \begin{aligned} 
\Qcircuit @C=1em @R=.7em @!R { 
& \prepareC{\sigma } & \qw \poloFantasmaCn{\rA} &\qw }  
\end{aligned} 
+  p_1  \begin{aligned} 
\Qcircuit @C=1em @R=.7em @!R {  
& \prepareC{\tau } & \qw \poloFantasmaCn{\rA} &\qw }  
\end{aligned}  
\end{align}  
where $p_0$ is a cancellative scalar. 
Let $\Sigma$ be an extension of $\sigma$ on system $\rE$, so that  	
\begin{align}
\begin{aligned} \Qcircuit @C=1em @R=.7em @!R {  
&   \multiprepareC{1}{\Sigma} & \qw \poloFantasmaCn{\rA} & \qw       & &=&   & \prepareC{\sigma } & \qw \poloFantasmaCn{\rA} &\qw  \\ 
& \pureghost{\Sigma} & \qw \poloFantasmaCn{\rE} & \measureD{e}} &  & &  &&  &   \end{aligned}
\end{align} 
for some deterministic effect $e$.  
Let $\eta$ be a deterministic state of system $\rE$ and let $\Theta$ be the extension of $\tau$ defined by $\Theta   :=  \tau  \otimes \eta$, so that   
\begin{align}
\begin{aligned} \Qcircuit @C=1em @R=.7em @!R {  
&   \multiprepareC{1}{\Theta} & \qw \poloFantasmaCn{\rA} & \qw       & &=&   & \prepareC{\tau } & \qw \poloFantasmaCn{\rA} &\qw  \\ 
& \pureghost{\Theta} & \qw \poloFantasmaCn{\rE} & \measureD{e}} &  & &  &&  &   \end{aligned}
\end{align} 
 Then, define the randomized preparation test $ {\bs \Delta }  : =  (\Delta_1, \Delta_2)$, with $\Delta_1  =  p_0\,  \Sigma$ and $\Delta_2  =  p_1\,   \Theta$ (the randomized preparation $\bs \Delta $ is a valid test due to the Randomization Assumption \ref{ass:random}).    The Displays Assumption \ref{ass:display} guarantees that the states  $ \Delta_1$ and $\Delta_2$ can be generated  as 
	\begin{align}
	\begin{aligned} 
	\Qcircuit @C=1em @R=.7em @!R {  
	& \multiprepareC{1}{\Delta_{i} } & \qw \poloFantasmaCn{\rA} &\qw   \\
		& \pureghost{\Delta_{i} } & \qw \poloFantasmaCn{\rE} &\qw   &\qquad   }  
		\end{aligned} =   
			\begin{aligned} \Qcircuit @C=1em @R=.7em @!R {&\qquad &  \multiprepareC{2}{\Gamma} & \qw \poloFantasmaCn{\rA} & \qw    
 \\ 
  &\qquad  & \pureghost{\Gamma} & \qw \poloFantasmaCn{\rE} & \qw\\
 &\qquad  & \pureghost{\Gamma} & \qw \poloFantasmaCn{\rF} & \measureD{f_i}  &  \qquad &\qquad    &\qquad  &\forall i\in  \{1,2\}}  \end{aligned}	\end{align} 
for some system $\mathrm{\rF}$, some  deterministic state $\Gamma $, and some measurement  $\su f  =  (f_1,f_2)$.  By construction,  $\Gamma$ is an extension of $\rho$: indeed, defining the deterministic effect $f :  =  f_1+f_2$, one has    
\begin{align}
\nonumber \begin{aligned} 
\Qcircuit @C=1em @R=.7em @!R {&  \multiprepareC{2}{\Gamma} & \qw \poloFantasmaCn{\rA} & \qw    
 \\ 
 & \pureghost{\Gamma} & \qw \poloFantasmaCn{\rE} & \measureD{e}\\
& \pureghost{\Gamma} & \qw \poloFantasmaCn{\rF} & \measureD{f} }  
\end{aligned}  \quad 
\nonumber &=  \sum_{i=1,2}
\begin{aligned} 
\Qcircuit @C=1em @R=.7em @!R {&  \multiprepareC{2}{\Gamma} & \qw \poloFantasmaCn{\rA} & \qw    
 \\ 
 & \pureghost{\Gamma} & \qw \poloFantasmaCn{\rE} & \measureD{e}\\
& \pureghost{\Gamma} & \qw \poloFantasmaCn{\rF} & \measureD{f_i} }   
 \end{aligned}  \\
 \nonumber  & \\
 \nonumber &  =  \sum_{i=1,2}
\begin{aligned} 
\Qcircuit @C=1em @R=.7em @!R {  
& \multiprepareC{1}{\Delta_{i} } & \qw \poloFantasmaCn{\rA} &\qw   \\
& \pureghost{\Delta_{i} } & \qw \poloFantasmaCn{\rE} &\measureD{e}  }  
\end{aligned}\\
\nonumber & \\
\nonumber &=
p_0  \,  \begin{aligned} 
\Qcircuit @C=1em @R=.7em @!R {  
& \multiprepareC{1}{\Sigma } & \qw \poloFantasmaCn{\rA} &\qw   \\
& \pureghost{\Sigma } & \qw \poloFantasmaCn{\rE} &\measureD{e}  }  
\end{aligned}  \\
\nonumber & \\
\nonumber & \quad		
+ p_1\,    \begin{aligned} 
\Qcircuit @C=1em @R=.7em @!R {  
& \multiprepareC{1}{\Theta } & \qw \poloFantasmaCn{\rA} &\qw   \\
& \pureghost{\Theta} & \qw \poloFantasmaCn{\rE} &\measureD{e}  }  
\end{aligned}\\
\nonumber & \\
\nonumber &   =  p_0\,       \begin{aligned} 
\Qcircuit @C=1em @R=.7em @!R {  
& \prepareC{\sigma } & \qw \poloFantasmaCn{\rA} &\qw }  
\end{aligned} 
+  p_1 \,    \begin{aligned} 
\Qcircuit @C=1em @R=.7em @!R {  
& \prepareC{\tau } & \qw \poloFantasmaCn{\rA} &\qw }  
\end{aligned}  \\
\nonumber &  \\
&  =  
\begin{aligned} 
\Qcircuit @C=1em @R=.7em @!R {  
& \prepareC{\rho } & \qw \poloFantasmaCn{\rA} &\qw }  
\end{aligned} 
   \, .
 \end{align}
 Now, suppose that $\map P$ and $\map P$ are two processes of type $\rA\to \rB$ and suppose that the relation
 \begin{align}
\begin{aligned} 
\Qcircuit @C=1em @R=.7em @!R {&  \multiprepareC{2}{\Gamma} & \qw \poloFantasmaCn{\rA} & \gate{\map P}  &  \qw \poloFantasmaCn{\rB} & \qw   
 \\ 
 & \pureghost{\Gamma} & \qw \poloFantasmaCn{\rE} & \qw  &\qw &\qw\\
& \pureghost{\Gamma} & \qw \poloFantasmaCn{\rF} & \qw &\qw &\qw }  
\end{aligned}
\quad=
\begin{aligned} 
\Qcircuit @C=1em @R=.7em @!R {&  \multiprepareC{2}{\Gamma} & \qw \poloFantasmaCn{\rA} & \gate{\map P'}  &  \qw \poloFantasmaCn{\rB} & \qw   
 \\ 
 & \pureghost{\Gamma} & \qw \poloFantasmaCn{\rE} & \qw  &\qw &\qw\\
& \pureghost{\Gamma} & \qw \poloFantasmaCn{\rF} & \qw &\qw &\qw }   
\end{aligned} 
\end{align}
holds. 
Applying the effect $f_1$ on both sides, we then obtain  
 \begin{align}
\begin{aligned} 
\Qcircuit @C=1em @R=.7em @!R {&  \multiprepareC{2}{\Gamma} & \qw \poloFantasmaCn{\rA} & \gate{\map P}  &  \qw \poloFantasmaCn{\rB} & \qw   
 \\ 
 & \pureghost{\Gamma} & \qw \poloFantasmaCn{\rE} & \qw  &\qw &\qw\\
& \pureghost{\Gamma} & \qw \poloFantasmaCn{\rF} & \qw &\qw &\measureD{f_1} }  
\end{aligned}
=
\begin{aligned} 
\Qcircuit @C=1em @R=.7em @!R {&  \multiprepareC{2}{\Gamma} & \qw \poloFantasmaCn{\rA} & \gate{\map P'}  &  \qw \poloFantasmaCn{\rB} & \qw   
 \\ 
 & \pureghost{\Gamma} & \qw \poloFantasmaCn{\rE} & \qw  &\qw &\qw\\
& \pureghost{\Gamma} & \qw \poloFantasmaCn{\rF} & \qw &\qw &\measureD{f_1} }   
\end{aligned}  ~,
\end{align}
and therefore  
\begin{align}
p_0\,  \begin{aligned} 
\Qcircuit @C=1em @R=.7em @!R {&  \multiprepareC{1}{\Sigma} & \qw \poloFantasmaCn{\rA} & \gate{\map P}  &  \qw \poloFantasmaCn{\rB} & \qw   
 \\ 
 & \pureghost{\Sigma} & \qw \poloFantasmaCn{\rE} & \qw  &\qw &\qw}  
\end{aligned}
=
p_0\,    \begin{aligned} 
\Qcircuit @C=1em @R=.7em @!R {&  \multiprepareC{1}{\Sigma} & \qw \poloFantasmaCn{\rA} & \gate{\map P'}  &  \qw \poloFantasmaCn{\rB} & \qw   
 \\ 
 & \pureghost{\Sigma} & \qw \poloFantasmaCn{\rE} & \qw  &\qw &\qw   }  
\end{aligned} \quad  .
\end{align}
Since $p_0$ is cancellative,  the two processes $\map P$ and $\map P'$ must coincide on $\Sigma$  (cf. Lemma \ref{lem:cancel}). Hence, one has $\Gamma\succeq_{\rA\to \rB}  \Sigma$.   \qed 
  
\section{Proof of Theorem \ref{theo:math}}\label{app:main}

{\bf  Proof of $2  \Longrightarrow 1$.}  Suppose that there exists a complete state, call it $\omega\in   \Det\St (\rA)$, such that the set of all extensions of $\omega$ has a maximum with respect to the tomographic ordering $\succeq_{\rA\to \rB}$.   Let $\Phi \in\Ext(\omega, \rR)$ be such maximum, for some auxiliary system $\rR$, and let   $\map P$ and $\map P'$ be two arbitrary processes of type $\rA \to \rB$.  By definition, the condition $(\map P\otimes \map I_{\rR})\Phi  =  (\map P' \otimes \map I_{\rR} )\,  \Phi $ implies the equivalence $\map P  \equiv_\omega  \map P'$.  In turn,   the condition  $\map P  \equiv_\omega  \map P'$ implies   $\map P= \map P'$  (Corollary  \ref{cor:complete}).    Hence, we obtained that the condition   $(\map P\otimes \map I_{\rR})\Phi  =  (\map P' \otimes \map I_{\rR} )\,  \Phi $ implies $\map P= \map P'$, meaning that $\Phi$ is dynamically faithful for processes of type $\rA\to \rB$.  \\

{\bf Proof of $1  \Longrightarrow 2$.}    Let $\Phi  \in  \St (\rA\otimes \rR)$  be a dynamically faithful state for processes of type $\rA\to \rB$.     Without loss of generality, the state $\Phi$ can be taken to be deterministic (Lemma \ref{lem:faithdet}).   Then,   let us pick a deterministic effect on system $\rR$, say $r\in  \Det\Eff (\rR)$, and define the state 
\begin{align}
\begin{aligned} \Qcircuit @C=1em @R=.7em @!R {  
&    \prepareC{\phi} & \qw \poloFantasmaCn{\rA} &\qw   &  &:= & &     \multiprepareC{1}{\Phi} & \qw \poloFantasmaCn{\rA} & \qw       \\ 
&  & &  &&  &&   \pureghost{\Phi} & \qw \poloFantasmaCn{\rR} & \measureD{r}  &.}   \end{aligned}
\end{align} 
By construction, $\Phi$ is an extension of $\phi$.   If $\phi$ is a complete state, this concludes the proof:  the complete state $\phi$ has an extension $\Phi$ that is a (global) maximum with respect to the tomographic ordering $\succeq_{\rA\to \rB}$.   

  If $\phi$ is not complete, we can pick a complete state $\sigma\in  \Det\St (\rA)$,   an arbitrary deterministic state  $\psi  \in  \Det\St(\rR)$, and, using the Randomizations Assumption \ref{ass:random}, we can define the source   $\bs \rho  =  (\rho_0, \rho_1)$   with  $\rho_0  :  = p_0\,  \Phi$ and $\rho_1  :  =     p_1\,  \sigma\otimes \psi$, where $(p_0,p_1)$ is a probability distribution such that $p_0$ and $p_1$ are both cancellative  (the existence of such probability distribution is guaranteed by the Coins Assumption \ref{ass:Coins}).  Note that the condition  $ ( \map P\otimes \map I_\rR)  \sim_{\bs \rho}\,  (  \map P'\otimes \map I_{\rR})$ implies  $p_0 \,   ( \map P\otimes \map I_\rR)    \Phi   =p_0 \,   (  \map P'\otimes \map I_{\rR})  \Phi$, which in turn implies   $  ( \map P\otimes \map I_\rR)    \Phi   =  (  \map P'\otimes \map I_{\rR})  \Phi$  (because $p_0$ is cancellative)  and $\map P  = \map P'$  (because $\Phi$ is dynamically faithful).  Now, let us denote the average state of the source $\bs \rho$  by $\rho:  =  \rho_0  + \rho_1   =  p_0 \,  \Phi  +  p_1\,  \sigma\otimes \psi$.  Note that one has 
	\begin{align}
\nonumber	\begin{aligned} \Qcircuit @C=1em @R=.7em @!R {  
	  &   \multiprepareC{1}{\rho} & \qw \poloFantasmaCn{\rA} & \qw  
 \\ 
	& \pureghost{\rho} & \qw \poloFantasmaCn{\rR} & \measureD{r}}  \end{aligned}	     &=  \quad   p_0 \,  \begin{aligned} \Qcircuit @C=1em @R=.7em @!R {  
	  &   \multiprepareC{1}{\Phi} & \qw \poloFantasmaCn{\rA} & \qw  
 \\ 
	& \pureghost{\Phi} & \qw \poloFantasmaCn{\rR} & \measureD{r}}  \end{aligned}	  \quad + \quad  p_1\,  \begin{aligned} \Qcircuit @C=1em @R=.7em @!R {  
	  &   \prepareC{\sigma} & \qw \poloFantasmaCn{\rA} & \qw  
 \\ 
	& \prepareC{\psi} & \qw \poloFantasmaCn{\rR} & \measureD{r}}  \end{aligned}	\\
\nonumber	&  \\
\nonumber 	&  =    \quad p_0  \,  \begin{aligned}   \Qcircuit @C=1em @R=.7em @!R {  
	  &   \prepareC{\phi} & \qw \poloFantasmaCn{\rA} & \qw}  \end{aligned}   \quad+ \quad  p_1\,    \begin{aligned} \Qcircuit @C=1em @R=.7em @!R {  
	  &   \prepareC{\sigma} & \qw \poloFantasmaCn{\rA} & \qw}  \end{aligned}\\
	  \nonumber &  \\
	  &  =:  \begin{aligned}   \Qcircuit @C=1em @R=.7em @!R {  
	  &   \prepareC{\omega} & \qw \poloFantasmaCn{\rA} & \qw}  \end{aligned}      \, ,
  \label{omega}\end{align}   
where  the state $\omega: = p_0 \,  \phi  +  p_1\,  \sigma$ is complete  (because  $\omega$ contains $\sigma$, and $\sigma$ is complete). 
Then, Lemma \ref{lem:intermediate} guarantees that there exists a deterministic state $\Gamma  \in  \Ext (\rho, \rE)$ such that the condition   $ ( \map P\otimes \map I_\rR\otimes \map I_\rE)  \Gamma   = ( \map P'\otimes \map I_\rR\otimes \map I_\rE)  \Gamma $ implies   $ ( \map P\otimes \map I_\rR)  \sim_{\bs \rho}\,  (  \map P'\otimes \map I_{\rR})$, and therefore $\map P= \map P'$.  Hence, the deterministic state $\Gamma$ is dynamically faithful.    Note that $\Gamma$ is an extension of $\rho$, which in turn is an extension of the complete state $\omega$ (by Eq. \eqref{omega}).  Summarizing, $\Gamma$ is an extension of a complete state and is a (global) maximum with respect to  the tomographic ordering $\succeq_{\rA\to \rB}$.  \qed

\section{Relations between Conclusive Teleportation and Universal Extension}\label{app:tele}

Let us  start by proving Proposition \ref{prop:tele}, which states that every teleportation state is  the universal extension of a  complete state.
\smallskip

{\bf Proof of Proposition \ref{prop:tele}.  }     Let $(E,F)$ be a binary measurement satisfying the teleportation equation  (\ref{teleportation}).  Let $\omega$ be a complete state of system $\rA$   (the existence of such state is guaranteed by   Assumption \ref{ass:complete}).   Let us define the state 
 \begin{align}\label{defchi}
 \begin{aligned} \Qcircuit @C=1em @R=.7em @!R { & \prepareC{\chi}& \qw \poloFantasmaCn{\rA}   & \qw }
 \end{aligned} \quad : = 
 \begin{aligned} \Qcircuit @C=1em @R=.7em @!R {
	 & \pureghost{\Phi} & \qw \poloFantasmaCn{\rA} &  \qw    \\
	 & \multiprepareC{-1}{\Phi} & \qw \poloFantasmaCn{\rR}   &\multimeasureD{1}{T}  \\
	  & \prepareC{\omega}& \qw \poloFantasmaCn{\rA}   &\ghost{T}      } 
\end{aligned}    \quad \, ,
 \end{align} 
 where  $T:  =  E+  F$  is  the deterministic effect obtained by coarse-graning over the binary measurement $(E,F)$.  
 By construction,  $\Phi$ is an extension of $\chi$.  Moreover, $\chi$ is complete.  To prove it, we pick an arbitrary deterministic state $\rho$ and write $\omega$ as $\omega  =  p_0 \, \rho +  p_1  \,  \tau$ for some state $\tau$ and for some  binary probability distribution $(p_0,p_1)$ such that $p_0$ is cancellative.   Inserting this expression in Eq.~\eqref{defchi}, we obtain
\begin{align}
\nonumber  \begin{aligned} \Qcircuit @C=1em @R=.7em @!R { & \prepareC{\chi}& \qw \poloFantasmaCn{\rA}   & \qw }
 \end{aligned}  &= 
  p_0 \, \begin{aligned} \Qcircuit @C=1em @R=.7em @!R {
	 & \pureghost{\Phi} & \qw \poloFantasmaCn{\rA} &  \qw    \\
	 & \multiprepareC{-1}{\Phi} & \qw \poloFantasmaCn{\rR}   &\multimeasureD{1}{T}  \\
	  & \prepareC{\rho}& \qw \poloFantasmaCn{\rA}   &\ghost{T}      } 
\end{aligned}   \quad  + \quad  p_1   \, \begin{aligned} \Qcircuit @C=1em @R=.7em @!R {
	 & \pureghost{\Phi} & \qw \poloFantasmaCn{\rA} &  \qw    \\
	 & \multiprepareC{-1}{\Phi} & \qw \poloFantasmaCn{\rR}   &\multimeasureD{1}{T}  \\
	  & \prepareC{\tau}& \qw \poloFantasmaCn{\rA}   &\ghost{T}      } 
\end{aligned} \\
\nonumber &  \\
  \nonumber  &  =  p_0 \, \begin{aligned} \Qcircuit @C=1em @R=.7em @!R {
	 & \pureghost{\Phi} & \qw \poloFantasmaCn{\rA} &  \qw    \\
	 & \multiprepareC{-1}{\Phi} & \qw \poloFantasmaCn{\rR}   &\multimeasureD{1}{E}  \\
	  & \prepareC{\rho}& \qw \poloFantasmaCn{\rA}   &\ghost{E}      } 
\end{aligned}  \quad  + \quad   p_0  \, \begin{aligned} \Qcircuit @C=1em @R=.7em @!R {
	 & \pureghost{\Phi} & \qw \poloFantasmaCn{\rA} &  \qw    \\
	 & \multiprepareC{-1}{\Phi} & \qw \poloFantasmaCn{\rR}   &\multimeasureD{1}{F}  \\
	  & \prepareC{\rho}& \qw \poloFantasmaCn{\rA}   &\ghost{F}      }   
\end{aligned}    \\
\nonumber &  \\
\nonumber &  \quad   +\quad   p_1   \, \begin{aligned} \Qcircuit @C=1em @R=.7em @!R {
	 & \pureghost{\Phi} & \qw \poloFantasmaCn{\rA} &  \qw    \\
	 & \multiprepareC{-1}{\Phi} & \qw \poloFantasmaCn{\rR}   &\multimeasureD{1}{T}  \\
	  & \prepareC{\tau}& \qw \poloFantasmaCn{\rA}   &\ghost{T}      } 
\end{aligned} \\
\nonumber &  \\
 \nonumber  &  =  p_0 \,  p_\rA  \,   \begin{aligned} \Qcircuit @C=1em @R=.7em @!R {
	 & \prepareC{\rho} & \qw \poloFantasmaCn{\rA} &  \qw  } 
\end{aligned}  \quad +\quad    p_0  \, \begin{aligned} \Qcircuit @C=1em @R=.7em @!R {
	 & \pureghost{\Phi} & \qw \poloFantasmaCn{\rA} &  \qw    \\
	 & \multiprepareC{-1}{\Phi} & \qw \poloFantasmaCn{\rR}   &\multimeasureD{1}{F}  \\
	  & \prepareC{\rho}& \qw \poloFantasmaCn{\rA}   &\ghost{F}      }   
\end{aligned}    \\
\nonumber &  \\
 &  \quad   +\quad     p_1  \, \begin{aligned} \Qcircuit @C=1em @R=.7em @!R {
	 & \pureghost{\Phi} & \qw \poloFantasmaCn{\rA} &  \qw    \\
	 & \multiprepareC{-1}{\Phi} & \qw \poloFantasmaCn{\rR}   &\multimeasureD{1}{T}  \\
	  & \prepareC{\tau}& \qw \poloFantasmaCn{\rA}   &\ghost{T}      } 
\end{aligned} 
 ~ \, .
 \end{align} 
Since $p_0$ and $p_\rA$ are both cancellative,  $p_0 p_\rA$ is cancellative, and therefore  $\chi$ contains $\rho$.  Since $\rho$ is arbitrary, we conclude that $\chi$ is complete.  

 It remains to prove that $\Phi$ is a universal extension of $\chi$.  This is easily done using the teleportation equation (\ref{teleportation}). Suppose that $\Gamma \in \Ext(\chi,  \rE)$ is another extension of $\chi$.  Then, one has 
 \begin{align}
\nonumber  p_A \,  
 \begin{aligned} \Qcircuit @C=1em @R=.7em @!R {
	 & \pureghost{\Gamma} & \qw \poloFantasmaCn{\rA} &  \qw    \\
	 & \multiprepareC{-1}{\Gamma} & \qw \poloFantasmaCn{\rE}   &\qw}   
\end{aligned}  
 \quad  & =  
\begin{aligned} \Qcircuit @C=1em @R=.7em @!R {
	 & \pureghost{\Phi} & \qw \poloFantasmaCn{\rA} &  \qw    \\
	 & \multiprepareC{-1}{\Phi} & \qw \poloFantasmaCn{\rR}   &\multimeasureD{1}{E}  \\
	  &\multiprepareC{1}{\Gamma} & \qw \poloFantasmaCn{\rA}   &\ghost{E}    \\
	  &\pureghost{\Gamma}  &  \qw  \poloFantasmaCn{\rE}   &\qw    } 
\end{aligned}    \\
\nonumber & \\
   &  =  \begin{aligned} \Qcircuit @C=1em @R=.7em @!R {
	 & \pureghost{\Phi} & \qw \poloFantasmaCn{\rA} &  \qw    &\qw &\qw   \\
	 & \multiprepareC{-1}{\Phi} & \qw \poloFantasmaCn{\rR}   &\gate{\map T}  &  \qw \poloFantasmaCn{\rE}  &\qw }   
\end{aligned}  ~,
\end{align}
 with
 \begin{align}
 \begin{aligned} \Qcircuit @C=1em @R=.7em @!R {
 & \qw \poloFantasmaCn{\rR} &\gate{\map T }  & \qw \poloFantasmaCn{\rE}  & \qw }   
\end{aligned}	     
\quad: = 
\begin{aligned} \Qcircuit @C=1em @R=.7em @!R {
	  & \qw & \qw \poloFantasmaCn{\rR}   &\multimeasureD{1}{E}  \\
	  &\multiprepareC{1}{\Gamma} & \qw \poloFantasmaCn{\rA}   &\ghost{E}    \\
	  &\pureghost{\Gamma}  &  \qw  \poloFantasmaCn{\rE}   &\qw    } 
\end{aligned}   
\end{align}
In summary, the extension $\Gamma$ is probabilistically obtained from the teleportation state $\Phi$. Since $\Gamma$ is arbitrary, $\Phi$ is a universal extension. 
\qed 

It is worth observing that the proof of Proposition \ref{prop:tele}  also implies a stronger result: in fact, {\em every}
 bipartite state of the composite  $\rA\otimes \rE$, with arbitrary $\rE$, can be probabilistically generated from the teleportation state $\Phi$ by means of a physical transformation.    In other words, the state $\Phi$ is preparationally faithful from system $\rR$ to system $\rE$, in the sense of Definition \ref{def:prepfaith}.

It is also worth noting that Conclusive Teleportation does not imply, in general, that every state has a universal extension.  
A condition that {\em would} imply the existence of   universal extensions for  generic  states $\rho$  is the existence of   conclusive teleportation protocols of the  form  
 \begin{align}\label{teleportationrho}
\begin{aligned} \Qcircuit @C=1em @R=.7em @!R {
	 & \pureghost{\Psi} & \qw \poloFantasmaCn{\rA} &  \qw    \\
	 & \multiprepareC{-1}{\Psi} & \qw \poloFantasmaCn{\rR}   &\multimeasureD{1}{F}  \\
	  & & \qw \poloFantasmaCn{\rA}   &\ghost{F}      } 
\end{aligned}    
\quad \equiv_\rho \quad p_\rho   \quad
\begin{aligned} \Qcircuit @C=1em @R=.7em @!R {
 & \qw \poloFantasmaCn{\rA} &\gate{\map I_\rA }  & \qw \poloFantasmaCn{\rA}  & \qw }   
\end{aligned}	     ~,
\end{align}
where $\Psi$ is an extension of $\rho$,  $F$ is a suitable effect, and $p_\rho$ is a cancellative scalar.

\begin{prop} 
If Eq. \eqref{teleportationrho}
 is satisfied, then the state $\Psi$ is a universal extension of $\rho$.   
 \end{prop}
\Proof   Suppose that $\Gamma \in \Ext (\rho, \rE)$ is another extension of $\rho$.  Then, Eq. \eqref{teleportationrho} implies the relation
 \begin{align}
 \nonumber p_\rho \,  
 \begin{aligned} \Qcircuit @C=1em @R=.7em @!R {
	 & \pureghost{\Gamma} & \qw \poloFantasmaCn{\rA} &  \qw    \\
	 & \multiprepareC{-1}{\Gamma} & \qw \poloFantasmaCn{\rE}   &\qw}   
\end{aligned}  
 \quad  & =  
\begin{aligned} \Qcircuit @C=1em @R=.7em @!R {
	 & \pureghost{\Psi} & \qw \poloFantasmaCn{\rA} &  \qw    \\
	 & \multiprepareC{-1}{\Psi} & \qw \poloFantasmaCn{\rR}   &\multimeasureD{1}{F}  \\
	  &\multiprepareC{1}{\Gamma} & \qw \poloFantasmaCn{\rA}   &\ghost{F}    \\
	  &\pureghost{\Gamma}  &  \qw  \poloFantasmaCn{\rE}   &\qw    } 
\end{aligned}    \\
 \nonumber  &  \\
   &  =  \begin{aligned} \Qcircuit @C=1em @R=.7em @!R {
	 & \pureghost{\Psi} & \qw \poloFantasmaCn{\rA} &  \qw    &\qw &\qw   \\
	 & \multiprepareC{-1}{\Psi} & \qw \poloFantasmaCn{\rR}   &\gate{\map T}  &  \qw \poloFantasmaCn{\rE}  &\qw }   
\end{aligned}  ~,
\end{align}
 with
 \begin{align}
 \begin{aligned} \Qcircuit @C=1em @R=.7em @!R {
 & \qw \poloFantasmaCn{\rR} &\gate{\map T }  & \qw \poloFantasmaCn{\rE}  & \qw }   
\end{aligned}	     
\quad: = 
\begin{aligned} \Qcircuit @C=1em @R=.7em @!R {
	  & \qw & \qw \poloFantasmaCn{\rR}   &\multimeasureD{1}{F}  \\
	  &\multiprepareC{1}{\Gamma} & \qw \poloFantasmaCn{\rA}   &\ghost{F}    \\
	  &\pureghost{\Gamma}  &  \qw  \poloFantasmaCn{\rE}   &\qw    } 
\end{aligned}   
\end{align}
In summary, the extension $\Gamma$ is probabilistically obtained from the teleportation state $\Psi$. Since $\Gamma$ is arbitrary, $\Psi$ is a universal extension. \qed 

\medskip  However, Eq. (\ref{teleportationrho}) does not appear as a particularly compelling requirement for general physical theories, and therefore we did not include it in our formulation of  Conclusive Teleportation.

\section{Proof of Proposition \ref{prop:puriisuniversal}}\label{app:puriisuniversal}

Let $\Phi \in \Pur\St (\rA\otimes \rE \otimes \rF)$ be a purification of the state $\Gamma$, with purifying system $\rE$. Since $\Gamma$ is an extension of $\rho$,  $\Phi$ is also a purification of $\rho$, with purifying system $\rE\otimes \rF$.    Now, pick an arbitrary deterministic pure state of $\rR$, say $\varphi \in  \Pur\St(\rR)$, and an arbitrary deterministic pure state of $\rE\otimes \rF$, say $\psi \in  \Pur\St (\rE \otimes \rF)$.  Then, Pure Product States guarantees that the two states $\Phi\otimes \varphi$ and $  \Psi \otimes \psi$ are both pure. By construction, both states  are  purifications of $\rho$.   Hence, the symmetry property of purifications implies that there exists a reversible process $\map U:   \rR \otimes \rE\otimes \rF \to \rE\otimes \rF \otimes \rR $ such that 
 \begin{align}
 	\begin{aligned} \Qcircuit @C=1em @R=.7em @!R {  
	  &   \multiprepareC{2}{\Phi} & \qw \poloFantasmaCn{\rA} & \qw     & &  &&      \multiprepareC{1}{\Psi} & \qw \poloFantasmaCn{\rA} & \qw   &\qw &\qw
 \\ 
	& \pureghost{\Phi} & \qw \poloFantasmaCn{\rE} & \qw   &\qquad&  = &  \qquad &  \pureghost{\Psi} & \qw \poloFantasmaCn{\rR} & \multigate{2}{\map U}   &  \qw  \poloFantasmaCn{\rE} &\qw \\
		& \pureghost{\Phi} & \qw \poloFantasmaCn{\rF} & \qw&&&&      \multiprepareC{1}{\psi} & \qw \poloFantasmaCn{\rE} & \ghost{\map U}   &\qw \poloFantasmaCn{\rF} &\qw\\
		& \prepareC{\varphi} & \qw \poloFantasmaCn{\rR} & \qw&&&&  \pureghost{\psi} & \qw \poloFantasmaCn{\rF} & \ghost{\map U}   &  \qw  \poloFantasmaCn{\rR} &\qw  
		} 
	 \end{aligned}
 \end{align}
 Applying the deterministic effect to system $\rF \otimes \rR$ we then obtain the condition  
 	 	\begin{align}
\nonumber &	\begin{aligned} \Qcircuit @C=1em @R=.7em @!R {  
	  &   \multiprepareC{1}{\Gamma} & \qw \poloFantasmaCn{\rA} & \qw      &  \qquad &  =  &  \qquad & 
	   \multiprepareC{2}{\Phi}  &   \qw \poloFantasmaCn{\rA} & \qw       
 \\ 
	& \pureghost{\Gamma} & \qw \poloFantasmaCn{\rE} & \qw &&&&
	\pureghost{\Phi}   &  \qw \poloFantasmaCn{\rE} & \qw \\
	&&&&&&&
	\pureghost{\Phi}  &\qw \poloFantasmaCn{\rF} & \measureD{u_{\rF}} \\
	&&&&&&&  
	\prepareC{\varphi}  &\qw \poloFantasmaCn{\rR} & \measureD{u_{\rR}} } 
	 \end{aligned}  &  \\
\nonumber  	 &   \\  
\nonumber  &  \qquad	  \qquad \quad   \quad~	\begin{aligned} \Qcircuit @C=1em @R=.7em @!R {  
	  = &\qquad  &      \multiprepareC{1}{\Psi} & \qw \poloFantasmaCn{\rA} & \qw   &\qw &\qw
 \\ 
	 &&  \pureghost{\Psi} & \qw \poloFantasmaCn{\rR} & \multigate{2}{\map U}   &  \qw  \poloFantasmaCn{\rE} &\qw \\
&&	  \multiprepareC{1}{\psi} & \qw \poloFantasmaCn{\rE} & \ghost{\map U}   &\qw \poloFantasmaCn{\rF} &\measureD{u_{\rm F}}\\
		&&  \pureghost{\psi} & \qw \poloFantasmaCn{\rF} & \ghost{\map U}   &  \qw  \poloFantasmaCn{\rR} &\measureD{u_{\rm R}}  
		} \end{aligned}  	
	    \\ \nonumber  &
	    \\   &
	      \qquad	  \qquad \quad   \quad~	\begin{aligned} \Qcircuit @C=1em @R=.7em @!R {  
	  = &\qquad  &      \multiprepareC{1}{\Psi} & \qw \poloFantasmaCn{\rA} & \qw   &\qw &\qw
 \\ 
	 &&  \pureghost{\Psi} & \qw \poloFantasmaCn{\rR} & \gate{\map T}   &  \qw  \poloFantasmaCn{\rE} &\qw } \end{aligned}  \quad ,
	\end{align} 
	having defined  
	\begin{align}
		\begin{aligned} \Qcircuit @C=1em @R=.7em @!R {  
	 & \qw \poloFantasmaCn{\rR} & \gate{\map T}   &  \qw  \poloFantasmaCn{\rE} &\qw  &  \qquad  &  :=  &\qquad  & &  \qw \poloFantasmaCn{\rR} & \multigate{2}{\map U}   &  \qw  \poloFantasmaCn{\rE} &\qw   \\ 
	  & &    &  &&  \quad  &    &\qquad  &  \multiprepareC{1}{\psi}   &  \qw \poloFantasmaCn{\rE} & \ghost{\map U}   &  \qw  \poloFantasmaCn{\rF} &\measureD{u_{\rF}}   \\ 
	   & &   &   & &    &    &\qquad  &    \pureghost{\psi}   &\qw \poloFantasmaCn{\rF} & \ghost{\map U}   &  \qw  \poloFantasmaCn{\rR} &\measureD{u_{\rR}}    &~. \\  
	  } \end{aligned} 
	\end{align}
	\qed

%


\end{document}